\documentclass[12pt]{article}
\usepackage{graphicx,amssymb,amsmath,tensor,empheq,color}

\usepackage{physics}

\usepackage{hyperref}
\hypersetup{colorlinks = true, linkcolor=black, citecolor=black, urlcolor=blue}

\makeatletter

\newlength{\fighskip} \fighskip=2pt
\newlength{\figvskip} \figvskip=3pt

\newcommand{\nn}{\nonumber}
\newcommand{\beq}{\begin{equation}}
\newcommand{\eeq}{\end{equation}}
\newcommand{\bea}{\begin{eqnarray}}
\newcommand{\eea}{\end{eqnarray}}

\newcommand{\1}{ \,  \raisebox{+0.14em}{{\hbox{{\rm \scriptsize ]}} \raisebox{-0.2em}{\kern-.8em\hbox{1}}}} \, }  % %

\title {Derivation of the two Schwarzians  effective action \\ for   the Sachdev-Ye-Kitaev  spectral form factor}

\author{Matteo A. Cardella\footnote{matteo.cardella@unimi.it }\\
\normalsize\it Dipartimento di Fisica, Universit\`a degli Studi di Milano and INFN, \\
\normalsize \it  via Celoria 16, 20133 Milan, Italy \\
}

\begin{document}

\setcounter{tocdepth}{2}

\maketitle
\begin{abstract}
The Sachdev-Ye-Kitaev  model  spectral form factor exhibits absence of information loss, in the form of a ramp and a plateau that are typical  in  random matrix theory.
 In   a large $N$ collective fields description,
 the ramp was reproduced by Saad, Shenker and Stanford \cite{Saad:2018bqo}
 by  replica symmetry breaking saddles.
We derive   a two sides  Schwarzians effective action for fluctuations around the ramp critical  saddles,
by computing  responses to a smeared version of  the two replica  kinetic kernel.
Our result confirms \cite{Saad:2018bqo}, where the form of the action was heuristically  guessed by  indirect
arguments  supported by numerical evidences.
\end{abstract}

\newpage
\tableofcontents

\section{Introduction}

The Bekenstein-Hawking \cite{Bekenstein:1974ax},\cite{Hawking:1974sw}  black hole  entropy   formula  in conjunction with the holographic principle \cite{tHooft:1993dmi},\cite{Susskind:1993mu}  suggest   that  a  region of  space
surrounded by a boundary surface  of finite area,   should  be described
   by a  finite dimensional Hilbert space. It is not known   how such an holographic description  works, except then in few notable cases. In the  context of   AdS/CFT \cite{Maldacena:1997re}, despite a large amounts of results,  several  relevant  questions related to black holes  cannot be cast in  quantitative  terms, due to the complexity of the boundary theory  at finite temperature.

   A manifestation of the black hole information problem in holographic models  \cite{Maldacena:2001kr}  appears  in the  behavior of  correlation functions  of boundary operators at very large separation times.
    Finiteness of the entropy  demands that  a   correlation function  cannot decay to zero at large times, since this would violate  quantum mechanics in the form of quantum information loss.  For the sake of illustration, let us consider a thermal  two point function for
a boundary operator $\hat{O}(t)$

\beq
G_{\beta}(t) =  \frac{1}{Z(\beta)} Tr \left(e^{-\frac{\beta}{2} H} \hat{O}(t) e^{-\frac{\beta}{2} H} \hat{O}(0)\right) = \frac{1}{Z(\beta)} \sum_{m, n}  e^{- \left(\frac{\beta}{2} - it\right)E_{n}}  e^{- \left(\frac{\beta}{2} + it\right)E_{m}} | \langle n | \hat{O} | m \rangle |^{2}, \label{2pts}
\eeq
where $Z(\beta)$ is the thermal  partition function, the dimension of the Hilbert space is finite and exponentially large in the entropy and we  omit to denote   dependence of $\hat{O}$ on spacial  coordinates.
At early times $G_{\beta}(t)$ decays exponentially in time as the effect of thermalization,  which in the bulk corresponds to  the    black hole quasi-normal modes relaxation.
 However, the  finite sum in (\ref{2pts})   cannot  not go  to zero at large times, instead it  keeps oscillating   with an amplitude exponentially small in the entropy  \cite{Dyson:2002pf},\cite{Goheer:2002vf},\cite{Barbon:2003aq}.
     On the other hand,  the above  requirement is not satisfied by a semiclassical   bulk theory. The reflection coefficient for an incoming wavepacket scattered  from the  black hole classical horizon diminishes with  boundary time, because of the increasing blueshift of the scattered particle, with an increasing penetrating power beyond the horizon of its wave function.   From this intuitive argument, one expects  the  two point function  to go to zero in the separation time infinite limit, which is indeed the case.  This process is described  by  black hole quasi normal modes  \cite{Horowitz:1999jd}.
On the other hand, one expects a departure from classical dynamics for the black hole horizon at times  exponentially large  in the entropy, when discreteness of the spectrum of the black hole microstates becomes relevant.  This effect  should conspire to reproduce the expected  erratic fluctuations in the two point function.
This is a non perturbative quantum gravity  effect that is  hard to be reproduced.
It is interesting to study this problem  in the   SYK model, as  its  collective  mode
 describes  the dynamics of a boundary graviton in a nearly two dimensional anti de Sitter spacetime ($nAdS_2$), which  accounts for  the dynamics of the horizon of  a low temperature nearly extremal black hole in four dimensions.

The above formulation  of the black hole information paradox in an holographic setup can be rephrased in terms  of a  somehow simpler and more universal correlation function   then (\ref{2pts}), which  does not contain   matrix elements of specific boundary operators $| \langle n | \hat{O} | m \rangle |^{2}$ but still exhibits the same phenomenon \cite{Papadodimas:2015xma}. This is the   so called spectral form factor

\beq
|Z(\beta, t)|^2  =  Z(\beta -it)Z(\beta + it) =  \sum_{m,n} e^{-\beta(E_n + E_m )} e^{it (E_n  - E_m ) }. \label{sum}
\eeq

While at short times  $|Z(\beta, t)|^2$  is of the order of the square of the thermal partition function  $Z(\beta)^{2}$, at very long times, of the order of the inverse of the  mean  energy level spacing,  the spectral form factor  reaches a  limiting value $Z(2 \beta)$ usually called the \emph{plateau}, due to a cancellation between the off diagonal contributions in the sum (\ref{sum}).
A direct computation at large times of the spectral form factor in theories with a well defined gravity dual such as super Yang Mills
is currently impossible. However, the Sachdev-Ye-Kitaev model (SYK)   \cite{kitaev:talk1}, \cite{kitaev:talk2},\cite{Sachdev-Ye93}, \cite{Georges}, \cite{Sachdev10}, \cite{Sachdev10b}, \cite{Sachdev} offers both a numerical and an analytic handle for studying  the spectral form factor. An accurate numerical analysis   of the SYK spectral form factor was  done in  \cite{Cotler2017}, (see also \cite{Garcia-Garcia:2016mno} for earlier related work), while a subsequent work \cite{Saad:2018bqo}    explains part of  the observed behavior in terms of  a large $N$    collective fields description\footnote{Recent works related to the SYK spectral form factor include \cite{Khramtsov:2020bvs},\cite{Winer:2020gdp},\cite{delCampo:2017bzr},\cite{Cardella:2021rij}.}.

  \vspace{.5 cm}

  SYK  is a statistical mechanics model    over an  ensemble  of quantum mechanical many body systems of Majorana fermions,  with random couplings of even order  $q \ge 4$   all  to all interactions.
A statistical average over an ensemble of quantum  systems is not equivalent to a single quantum mechanical model,
a fact that may cast doubts upon using SYK for discussing delicate issues like unitarity   in a black hole holographic description. Yet, there are qualitative and quantitative features that survive the ensemble  disorder average that make SYK  an interesting playground for discussing certain  quantum gravity issues related to black holes\footnote{There are also the so called colored tensor models, quantum mechanical models without disorder  that exhibit the same diagrammatic as SYK  \cite{Witten:2016iux}.}.

SYK  has a  collective fields description that in the large $N$ limit   at low temperatures/strong coupling  exhibits an interesting quasi conformal behavior,  with  aspects of a gravity dual.
In particular, as an  holographic  model for nearly extremal  black holes
it provides   an interesting arena for  testing   various ideas and proposals and sharpen open problems.
One of its attractive features   is a dominant   soft mode dynamics at low temperatures  \cite{kitaev:talk1},\cite{kitaev:talk2},\cite{Maldacena:2016hyu},\cite{Jevicki:2016ito},   that from a bulk perspective encodes the full black hole gravitational backreaction \cite{Maldacena:2016upp},\cite{Kitaev:2017awl},\cite{Kitaev:2018wpr},\cite{Maldacena:2017axo},   and,  under certain circumstances,  allows for  a   quantum mechanical  description   of  a  black hole interior \cite{Maldacena:2017axo},\cite{Kourkoulou:2017zaj},\cite{Brustein:2018fkr},\cite{Almheiri:2017fbd},\cite{Almheiri:2018ijj},\cite{Almheiri:2018xdw}.      One of the  exciting recent results, that is understood also from the SYK perspective,   is a   gravitational  description of the quantum teleportation protocol  \cite{Maldacena:2017axo},\cite{Maldacena:2018lmt},\cite{Susskind:2017nto}.
This  is based on  recent observations  on how certain   double trace deformations   that violate the average null energy  condition  make a wormhole temporarily traversable \cite{Gao:2016bin}. In the teleportation protocol those   double trace deformations   implement  the transmission of classical instruction for  the quantum protocol between the two boundaries through external space-time   and   make   the  wormhole connecting them  temporarily traversable.

\begin{figure}
  \includegraphics[width= 12 cm]{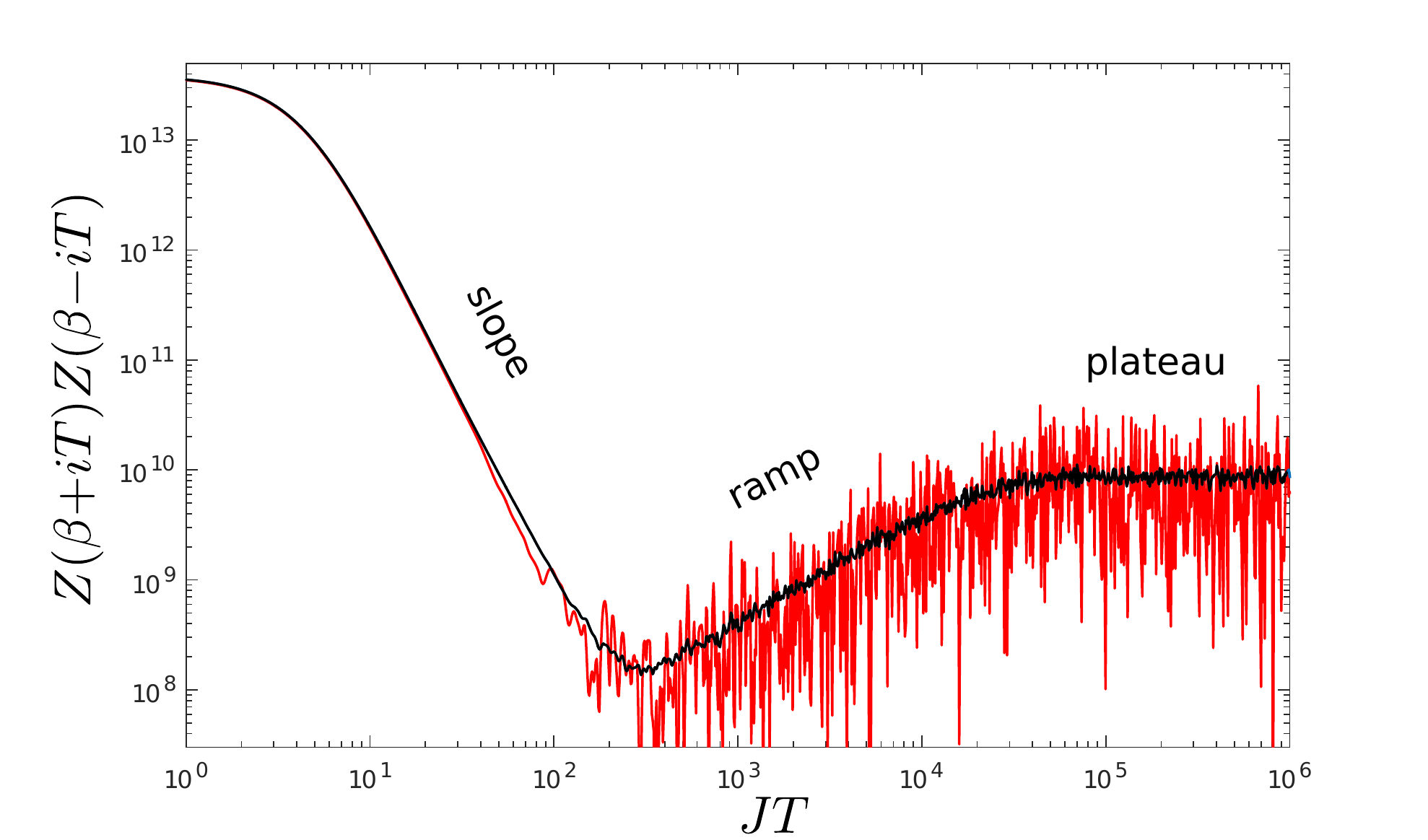}
  \caption{: A log-log plot \cite{Cotler2017} of the SYK spectral form factor for $q = 4$, $N = 34$, $\beta J = 5$. In red the time plot for a single realization of the disorder,  wild erratic oscillations are quite visible in  the ramp and the plateau parts of the plot.
 Oscillations are washed out by  disorder  average over a large enough number of samples without tough throwing  away the main trend.  Statistical average (in black) still exhibits an interested  lack of information loss in the form of a  random matrix universality. The  black   plot refers  to numerics from  an ensemble of   90 samples \cite{Cotler2017}.}
  \label{ZZbar}
\end{figure}

Another interesting feature is that SYK exhibits quantum chaotic behavior both at short and long times scales.
At short times scales,    out of time ordered (OTO) four point  thermal  correlators  involving a perturbation on  a typical  operator     saturate  \cite{kitaev:talk1},\cite{kitaev:talk2},\cite{Gu:2018jsv}  a universal  chaos bound \cite{Maldacena:2015waa} for   their  Liapunov exponent. In AdS/CFT,   saturation of the chaos bound in the boundary  is understood  \cite{Shenker:2013pqa}  by  a corresponding    bulk   near horizon scattering  dynamics  that involves     Dray and 't Hooft shock waves  \cite{Dray:1984ha}. The requirement for the  boundary theory   to be a  fast scrambler \cite{Sekino:2008he}   avoids black hole  quantum cloning in certain gedanken experiments \cite{Susskind:1993mu},\cite{Hayden:2007cs}. The fast scrambler property  is satisfied by  boundary theories that are $q$-local quantum many body systems with all to all interactions \cite{Sekino:2008he}. In the  SYK case   there are no shock waves in the two dimensional bulk and  the only bulk gravitational  degrees of freedom   are boundary modes  that undergo   classical chaotic dynamics by moving in  hyperbolic space \cite{Maldacena:2016upp},\cite{Maldacena:2017axo},\cite{Brown:2016wib},\cite{Lin:2019qwu}.
On the other hand,  SYK  exhibits also  chaotic behavior at large time scales\footnote{ controlled  by times scales exponentially large in $N$},  related  to the   fine  details of  correlations among  its energy eigenvalues.  The connected part of the  two point correlation function  for the SYK spectral density  controls the large time behavior of  the spectral form factor,  a quantity of interest    to diagnose  the existence of  quantum  information loss.
Numerics on the SYK spectral form factor  exhibit   the emergence at long time scales\footnote{Meaning exponential in $N$ time scales.} of  typical behaviour of  quantum  chaotic systems, in particular   random matrix theory (RMT) universality.
We refer to \cite{Cotler2017}   for  relevant numerical  plots and typical  values of the SYK spectral form factor, here we would like to  mention the
main qualitative features of the time behavior of the spectral form factor.
The plot  in figure \ref{ZZbar},    starts at early times  with a decaying phase  from its initial value $Z(\beta)^2$ with a characteristic decreasing  power law up to a minimum value. This decaying part of the plot is called the \emph{slope} and the minimum value is called  the \emph{dip}, which occurs at
 $t_{dip} \sim  e^{N / 2}$.  After the  dip, the SFF  starts a linear rising behavior  called the \emph{ramp}.  The ramp occurs up  to a saturation time  $t_{plateu} \sim e^{N}$  where $|Z(\beta, T)|^2$ reaches the  limiting value $Z(2\beta)$.
The ramp and the  plateau  are typical of quantum  chaotic systems  in particular in random matrix theory(RMT).
The emergence of  RMT universality in itself  is not surprising from a Hilbert space perspective, the interest, motivated by the black hole information problem, is to understand  the behavior of the spectral form factor  by using a large $N$ collective fields description.
This problem was discussed   in   \cite{Saad:2018bqo} for the   SYK model  and   by the same authors in  \cite{Saad:2019lba}  for Jackiw Teitelboim (JT) gravity \cite{Teitelboim:1983ux},\cite{Jackiw:1984je}, where a non perturbative completion  of JT gravity in terms of RMT is proposed.  In \cite{Stanford:2019vob}
the relations between JT gravity and RMT are extended to the case where  the boundary theory  has time-reversal symmetry and  have fermions with or without supersymmetry.

An interesting point that emerges from the numerics on the SYK spectral form factor  is that for a  single  realization of the disorder,  the time plot  exhibits  erratic oscillations  in its ramp and plateau regions,(figure \ref{ZZbar}).  Oscillations are  washed  out either by averaging SYK  over a large enough ensemble   or by taking a suitable  time average of the spectral form factor.
Although part of the information that corresponds to the erratic wild oscillations is washed out by the disorder average,  the main
trend that does not exhibit information loss is still present after ensemble average. This provides an interest for  a  quantitative study of the SYK spectral form factor   in terms of   large $N$ collective fields, in  relation to the black hole information paradox.

It is easy to check that the  connected part of the disorder averaged  two point function for the analytically continued thermal partition function

\bea
\langle Z(\beta + iT) Z(\beta - iT)  \rangle_{c} &\equiv&
\langle Z(\beta + iT) Z(\beta - iT)  \rangle -  \langle Z(\beta + iT) \rangle \langle  Z(\beta - iT)  \rangle  \nn \\
 &=& \int dE dE' \,  \langle \rho(E) \rho(E') \rangle_{c} \, e^{- \beta (E + E')} e^{iT(E -E')}  \label{Zc}
\eea

exhibits a linear ramp for  a contribution of the form  $\langle \rho(E) \rho(E') \rangle_{c} \sim  \frac{1}{(E - E')^{2}} $ in the connected component of the two point correlator for the  spectral  density of states $\rho(E)$. In the above expression brackets in the l.h.s.   denote disorder average, while brackets  in the r.h.s. denotes statistical correlation.  It is indeed  the   disorder average over  a statistical  ensemble that provides a non zero  connected component $\langle Z(\beta + iT) Z(\beta - iT)  \rangle_{c}$  to the analytically continued partition function.
In a  single quantum mechanical model the partition function is just a number and  there would be no two points  connected contributions whatsoever.
The above relation indicates that in order to study the connected component of the spectral form factor, in terms of a functional integral in  collective fields,
one should look to a two replica system. In fact, disorder average creates interactions between distinct   replica, which leads to a  connected component for the spectral form factor.
Another  indication that the non decaying contributions for the SYK spectral form factor  may be   understood by   a two replica systems may come   from
 the  ER = EPR conjecture \cite{Maldacena:2013xja},\cite{Maldacena:2001kr},\cite{Israel:1976ur}. Indeed, it turns out \cite{Saad:2018bqo} that  the family of   off replica diagonal   saddle points responsible for the SYK  ramp are  obtained with good approximation from a sum over images of   SYK correlators on the    thermofield double state. On the other hand, from the JT gravity
side, the ramp  is reproduced  by a JT  double trumpet  instanton   \cite{Saad:2019lba}, a connected Euclidean baby universe  that connects   two identical Euclidean  black holes at the same temperature.

\vspace{.5 cm}

In this paper  we derive  a time reparametrization  soft modes effective action that corrects the SYK critical (replica non-diagonal) saddles  of the spectral form factor (\ref{Zc}). This effective action governs the late times dynamics for the spectral form factor.
 We follow a method inspired by  \cite{Kitaev:2017awl} to study enhanced responses to the conformal breaking kinetic operator in SYK.
 This two sides  Schwarzian effective action  that we obtain,  appear also in \cite{Saad:2018bqo}.
In that work, the form of the effective action  was guessed  by  indirect reasoning, supported  by  numerical evidence \cite{DS}.
On the other hand,  we carry on a constructive analysis by outlining  the  dominant quantum effects that correct the spectral form factor correlation functions  at late times. Our results as a byproduct allow  to obtain   an explicit derivation  of the late times effective action.
A point of interest of the analysis carried here is a direct observation of when and how  non Schwarzian contributions
to the effective action become  ineffective  for the spectral form factor. This kind of analysis has also  some interesting points of contact with the recently appeared work  \cite{Zhang:2020jhn}.

\vspace{.5 cm}

   The organization of the paper is the following, in section \ref{sec1}  we go through   the construction of a  two replica functional integral for the   SYK spectral form factor, by going into details in the derivation of the replica non-diagonal conformal saddles.
In section \ref{sec2} we compute a large $N$ approximation for the spectral form factor by steepest descent method, around conformal  replica  non-diagonal saddles. We then  analyse the effects of a smeared version of the two replica   kinetic operator and single out that a  specific projection  of the kinetic operator   dominates the late  dynamics of the spectral form factor.
 In section $4$ we compute the soft modes effective action for the spectral form factor that governs the late times dynamics.
 In appendix, we review  the method presented  in  \cite{Kitaev:2017awl}  for the case  of  regular SYK.

\vspace{.5 cm}
\section{ Two replica path integral representation for the connected part of the spectral form factor} \label{sec1}

 The Sachdev-Ye-Kitaev model  (SYK) \cite{Sachdev-Ye93},\cite{Georges},\cite{Sachdev10},\cite{Sachdev10b},\cite{Sachdev},\cite{kitaev:talk1},\cite{kitaev:talk2} is a  statistical mechanics model  over an ensemble of many body systems given  by   $N$ (even) Majorana fermions $\psi_i = \psi_i ^{\dag}$

\beq
\{ \psi_i , \psi_j  \} = \delta_{ij},
\eeq

on a complete  hypergraph  of even   order $q \ge 4$.  Each system of  the ensemble has   Hamiltonian

\beq
H =  \frac{(i)^{\frac{q}{2}}}{q!} \sum_{i_1 \dots i_q} J_{i_1 \dots i_q} \psi_{i_1} \dots  \psi_{i_q}, \label{H}
\eeq

where the  couplings  $J_{i_1, \dots, i_q}$  are random variables, independently taken  from a    Gaussian probability distribution     with  zero mean $ \langle J_{i_1, \dots, i_q} \rangle  = 0$   and  variance  $ \langle J_{i_1, \dots, i_q}^2 \rangle  =   J^2/ N^{q-1}$. The parameter  $J$  fixes the characteristic  energy scales of the
model. In the large $N$ limit,  at  high temperatures/weak coupling    $\beta J << 1$,  the model is asymptotically free,  while   at low temperatures/strong coupling  $\beta J >>  1$, SYK develops an interesting quasi-conformal dynamics. Conformal symmetry is  slightly  explicitly  broken   by   a dominant   $O\left(\frac{\beta J}{N}\right)$ corrections,   with  leading contribution   coming    from a local  Schwarzian effective action.
 The same  pattern of symmetry breaking and a Schwarzian effective action
 occur in   a certain  limit  of two  dimensional  Jackiw-Teitelboim (JT) gravity. Moreover,  JT gravity
 describes the classical dynamics of the  near horizon region of a low temperature nearly extremal black hole.
 In the following we mainly  focus on the $q=4$ case,  extensions of the results for generic even $q$ are usually straightforward\footnote{For a nice review of the SYK model and its relations to JT gravity see   \cite{Sarosi:2017ykf}.}.

We consider the following connected   contribution  to  the SYK  disorder averaged analytically continued  thermal partition function two point correlator

\beq
\langle Z(\beta - iT )Z(\beta + iT) \rangle_{c} \equiv \langle Z(\beta - iT )Z(\beta + iT) \rangle  -   \langle Z(\beta - iT ) \rangle \langle Z(\beta + iT) \rangle, \label{Zconn}
\eeq

where brackets denote disorder average. As already  remarked in the introduction, in order  to have a non vanishing connected component we shall look for a system of two replicas on which to perform disorder average. In fact, disorder average creates an interaction among  non interacting replica.
 By following \cite{Saad:2018bqo}  we construct a functional integral representation for
$\langle Z(\beta - iT )Z(\beta + iT) \rangle_{c}$ in terms of two copies or replica of SYK,  $SYK_{L}$ and  $SYK_{R}$.
 We consider the following two replica  representation for the spectral form factor for one particular realization of the disorder

\beq
Z(\beta - iT)Z(\beta + iT)  = Tr \left( e^{- (\beta - iT)H_L } e^{- (\beta + iT)H_R }  \right) = Tr \left( e^{-T \left( J_L H_L  + J_R H_R  \right)}   \right) \label{Z2}
\eeq
where
  \beq
H_{I} =  - \frac{1}{4!} \sum_{i,j,k,l} J_{ijkl} \psi_{i}^{I}\psi_{j}^{I} \psi_{k}^{I}\psi_{l}^{I},   \qquad  I=L,R
\eeq

and
\beq
J_L = \frac{\beta}{T} - i,    \qquad  J_R = \frac{\beta}{T} + i =  J_{L}^{*}. \label{trickt2}
\eeq

By  disorder  averaging    (\ref{Z2})   one finds  the   following path integral representation

\bea
&& \langle Z(\beta + iT) Z(\beta - iT) \rangle_{c}  =  \nn \\ &&  \int \mathcal{D} \psi_i^{I} \exp \left( - \int_{0}^{T} d \tau \psi_{i}^{I} (\tau) \partial_{\tau} \psi_{i}^{I} (\tau)  +   \frac{N J^2 J_{I}J_{J}}{8} \int_{0}^{T} d \tau d \tau'  \left( \frac{1}{N} \psi_{i}^{I}(\tau) \psi_{i}^{J}(\tau')\right)^{4} \right), \nn \\ \label{Zaveraged2}
\eea
where replica indexes $I,J = L,R $ are summed up.

The above connected component has  to  be contrasted with the disconnected part of the two point function
   $\langle Z_{SYK}(\beta - iT ) \rangle \langle  Z_{SYK}(\beta + iT) \rangle $, obtained by the  standard SYK disorder averaged thermal partition function   $\langle Z_{SYK}(\beta) \rangle $ by analytic continuation  $\beta \rightarrow \beta  \pm iT$  .      $\langle  Z_{SYK}(\beta) \rangle$  can be computed at full quantum level in the Schwarzian approximation,
 since  the  Schwarzian path integral over   soft modes  is  one loop exact \cite{Stanford:2017thb}.
 This  same result for    $\langle Z_{SYK}(\beta) \rangle$     can also  be obtained    by solving a  quantum mechanical  problem  for a particle scattered by a    Liouville potential \cite{Bagrets:2016cdf},\cite{Bagrets:2017pwq}.
 Accurate methods for computing Schwarzian amplitudes at full quantum level are developed in \cite{Mertens:2017mtv},\cite{Mertens:2018fds}.
 The full quantum answer   is given by

 \beq
 \langle Z_{SYK}(\beta) \rangle \sim \frac{e^{+ \frac{\alpha_{S}}{\mathcal{J}} \frac{2\pi^2}{\beta}}}{(\mathcal{J}\beta)^{3/2}},
 \eeq

 which  gives the following disconnected contribution to the spectral form factor

 \beq
  \langle Z_{SYK}(\beta + iT ) \rangle  \langle Z_{SYK}(\beta - iT) \rangle   \sim \frac{e^{\frac{\alpha_{S}}{\mathcal{J}} \frac{4\pi^2 \beta}{\beta^2 + T^2 }}}{(\mathcal{J}^{3/2} (\beta^{2} + T^{2})^{3/2}}. \label{Zdiscn}
 \eeq

 This disconnected  contribution  reproduces accurately the decaying slope of  the spectral form factor in figure  \ref{ZZbar}.
 It is not surprising that this contribution manifests  information loss,  since it is given   by  a product of  analytic continuations of the SYK  thermal partition function. From the two replica system perspective, it is the result of a replica diagonal saddle plus the one loop determinant from fluctuations, that together give the full quantum answer.
  In contrast,  the connected  contribution from off diagonal  replica saddle is somehow related to  a   purification of the thermal density matrix,   obtained by  doubling the system. Indeed it turns out that the connected saddles can be written in terms of
 an antisymmetrized version of the thermofield double correlators \cite{Saad:2018bqo}, (see eq. (\ref{correlSFF2})  and related discussion).

Concerning the   functional integral in (\ref{Zaveraged2}),
before switching from the   representation in terms of  Majorana fermions to a  more convenient  description  in terms of $O(N)$ singlets collective fields,
let us notice that in any regime where the effects of the bilocal kinetic operator in (\ref{Zaveraged2})

\beq
 \hat{\sigma}_{IJ}(\tau, \tau') \equiv  \delta_{IJ} \delta(\tau -  \tau')  \partial_{\tau} \label{kin}
 \eeq

 can be neglected, the eight fermions interaction vertex in (\ref{Zaveraged2})  is invariant under the  following  transformation

\beq
\Psi_{i}^{I}(\tau_{I}) \rightarrow \left(f_{I}'(\tau_{I})\right)^{\Delta}\Psi_{i}^{I}(f_{I}(\tau_{I})),  \qquad  \Delta = \frac{1}{4},  \qquad I=L,R, \label{invFermions}
\eeq

 for two independent  time reparametrization  diffeomorphisms, (aka left and right soft modes),  $f_{I}(\tau_{I})$, $I=L,R$. In fact, it can be checked easily that  this  transformation   corresponds
to a change of integration variables in the double integral  interaction term in (\ref{Zaveraged2}).
Therefore in any regime where $\hat{\sigma}_{IJ}(\tau, \tau')$ (\ref{kin})  can be neglected, the system develops  the   time reparametrization symmetry (\ref{invFermions}), where fermions are primary fields of weight $\Delta = \frac{1}{4}$.
Let us notice also that  fermions  appear in the interaction term in (\ref{Zaveraged2}) as the following  $O(N)$ singlet  collective field

\beq
G_{IJ}(\tau,\tau') =  \frac{1}{N}\sum_{i=1}^{N} \psi_{i}^{I}(\tau) \psi_{i}^{J}(\tau'). \label{Gij}
\eeq

As a consequence of (\ref{invFermions}), in any  regime where the kinetic operator  $\hat{\sigma}_{IJ}(\tau, \tau')$ (\ref{kin}) can be ignored,
the action is invariant under the following reparametrization of the bilocal field

\beq
G_{IJ}(\tau_{I},\tau_{J})  \rightarrow  \left(f_{I}'(\tau_{I}) f_{J}'(\tau_{J})\right)^{\Delta} G_{IJ}( f_{I}(\tau_{I}) f_{J}(\tau_{J})),  \qquad \Delta = \frac{1}{4}, \qquad I=L,R. \label{Gtransf}
\eeq

\vspace{.3 cm}

In order to  study  $\langle Z(\beta + iT) Z(\beta - iT) \rangle_{c}$ in the  large $N$  limit, it is convenient  to recast  the functional integral  (\ref{Zaveraged2})     in terms of the collective field  $G_{IJ}(\tau,\tau')$  (\ref{Gij}) and a corresponding  Lagrangian multiplier
$\Sigma_{IJ}(\tau,\tau')$, by integrating out the Majorana fermion fields.  This  is  achieved by  inserting in the path integral (\ref{Zaveraged2}) the identity

\bea
1 &=& \int \mathcal{D}G_{IJ} \, \,   \delta   \left(G_{IJ}(\tau,\tau') -  \frac{1}{N}\sum_{i=1}^{N} \Psi_{i}^{I}(\tau) \Psi_{i}^{J}(\tau')  \right) \nn \\
&=&   \int \mathcal{D}G_{IJ}   \int \mathcal{D}\Sigma_{IJ} e^{\Sigma_{IJ}(\tau, \tau') \left(G_{IJ}(\tau, \tau') -  \frac{1}{N}\sum_{i=1}^{N} \Psi_{i}^{I}(\tau) \Psi_{i}^{J}(\tau')  \right)},
\eea

where integration over  $\Sigma_{IJ}$  is performed along an  imaginary direction in field space.
By   integrating  out  fermions  one finds

\beq
 \langle Z(\beta - iT) Z(\beta + iT) \rangle_{c}  = \int \mathcal{D}G_{ij}(\tau, \tau') \mathcal{D} \Sigma_{ij}(\tau, \tau') e^{-  I(G_{ij},\Sigma_{ij})}, \label{functintegral}
\eeq

where
\bea
  \frac{I(G_{ij},\Sigma_{ij}) }{N} = -  \frac{1}{2} \log \det \left( \delta(\tau -  \tau') \delta_{ij} \partial_{\tau} - \Sigma_{ij}(\tau, \tau')  \right)  + \nn \\  \frac{1}{2} \sum_{i,j} \int_{0}^{T} d \tau d \tau'  \left( \Sigma_{ij}(\tau, \tau' ) G_{ij}(\tau, \tau') - \frac{J^2 J_{ij} }{4}  G_{ij}(\tau, \tau')^4 \right). \label{I}
\eea

Notice that for  notational convenience  from now on we  use  lowercase indexes to denote left and right replicas  entries, by switching our previous  notation $I =L,R$ to  $i= L,R$. This should not be source of confusion, since  fermions have been integrated out.
 We also use the notation $J_{ij} = J_{i}J_{i}$ where

\beq
J_{ij} =
\begin{pmatrix}
 J_{L}J_{L} & J_{L}J_{R} \\
 J_{R}J_{L} & J_{R}J_{R}
\end{pmatrix}
=
\begin{pmatrix}
 \left( \frac{\beta}{T} + i \right)^2 &  \frac{\beta^2}{T^2}  + 1   \\
  \frac{\beta^2}{T^2}  + 1  &    \left( \frac{\beta}{T} - i \right)^2
\end{pmatrix}, \label{Jmatrix}
\eeq

which follows from the definitions  (\ref{trickt2}).

The two replica action (\ref{I}) gives the following saddle point Schwinger Dyson equations. A variation with respect $\Sigma_{ij}$

\beq
\frac{\delta I}{ \delta \Sigma_{ij}} = 0,
\eeq

 gives

\beq
\left( \delta(\tau - \tau'') \delta_{ik} \partial_{\tau}      -  \Sigma_{ik}(\tau, \tau'')\right) * G_{kj}(\tau'' , \tau' ) = - \delta_{ij}\delta(\tau -  \tau'), \label{SD1}
\eeq

where $*$ is the  convolution product

\beq
f * g (x) \equiv \int d y f(x - y)g(y).
\eeq

While a variation with respect to $G_{ij}$

\beq
\frac{\delta I}{ \delta G_{ij}}  = 0,
\eeq

gives

\beq
 \Sigma_{ij}(\tau, \tau') = J^{2}J_{ij} G_{ij}^3(\tau, \tau'). \label{SD2}
\eeq

 At strong coupling/low temperatures  $\beta J >> 1$,  an almost   conformal  regime emerges. It can be observed by considering
 the  Fourier transform of the saddle point equation  (\ref{SD1})

\beq
\left( - i \omega \delta_{ik}     -  \Sigma_{ik}(\omega) \right)  G_{kj}(\omega ) = - \delta_{ij}. \label{seq}
\eeq
By  taking the following  ansatz for the self energy $\Sigma_{ij}(\omega) = \Phi_{ij} \sqrt{ J \omega}$, where  $\Phi_{ij}$  is a constant invertible  matrix,  in the low energy  regime  $\omega  <<  J$,  the $i\omega$ term in (\ref{seq})  can be neglected w.r.t.  the  second one.   The conformal Schwinger Dyson equation therefore  reads

\beq
 \Sigma_{ik}(\omega)  G_{kj}(\omega ) = \delta_{ij}.
\eeq

The above equation together with the ansatz we made  for  $\Sigma_{ij} (\omega)$   gives

\beq
 G_{ij}(\omega) =  \frac{\Phi_{ij}^{-1}}{\sqrt{J \omega}}.   \label{GPhi}
\eeq

This latter relation when casted into the other saddle point equation  eq. (\ref{SD2}) leads to the following equation for the constant invertible matrix  $\Phi_{ij}$

\beq
\Phi_{ij} =   J_{ij} (\Phi_{ij}^{-1})^{3}. \label{Phi}
\eeq

eq. (\ref{Phi})  is solved by

\beq
\Phi_{ij}^{-1} =
\frac{1}{2^{1/4}}  \begin{pmatrix}
 \left( \frac{\beta}{T} + i \right)^{- 1/2} &  i \left( \frac{\beta^2}{T^2}  + 1 \right)^{- 1/4}  \\
   i \left( \frac{\beta^2}{T^2}  + 1 \right)^{- 1/4} &   \left( \frac{\beta}{T} -  i \right)^{- 1/2}
\end{pmatrix}, \label{Phisol}
\eeq

By casting (\ref{Phisol})  into  (\ref{GPhi}) one finds
 the following family of  complex  saddles for the SYK spectral form factor in the conformal limit
 \cite{Saad:2018bqo}

\bea
G_{LL}^{\beta_{aux}}(t) &=&  \frac{1}{(4\pi)^{1/4}\sqrt{J}} \left( \frac{\beta - iT}{\beta + iT} \right)^{1/4}  \frac{ \text{sgn}(t)}{\left[ \frac{\beta_{aux}}{\pi}\left| \sinh \left(\frac{\pi t}{\beta_{aux}} \sqrt{1 + \frac{\beta^2}{T^2}} \right)   \right| \right]^{1/2}}, \nn \\
G_{LR}^{\beta_{aux}}(t) &=& G_{RL}^{\beta_{aux}}(t) =  \frac{1}{(4\pi)^{1/4}\sqrt{J}}   \frac{i}{\left[ \frac{\beta_{aux}}{\pi} \cosh \left(\frac{\pi (t + \Delta)}{\beta_{aux}} \sqrt{1 + \frac{\beta^2}{T^2}} \right)   \right]^{1/2}}, \nn \\
G_{RL}^{\beta_{aux}}(t) &=&  (G_{LR}^{\beta_{aux}}(t))^{*}. \nn \\
 \label{correlSFF}
\eea
By following  the same convention used in \cite{Saad:2018bqo}, we  label  the effective auxiliary temperature
in the correlators by  $\tilde{\beta}_{aux} =  \frac{\beta_{aux}}{ \sqrt{1 + \frac{\beta^2}{T^2}}}$. The phase space of the family of nondiagonal backgrounds is two dimensional,  labelled  by   $\beta_{aux} > 0$
and  by the compact coordinate  $\Delta \in [-T, T]$. $\Delta$  is a  relative time shift between the two holographic boundaries clocks.

Let us notice that the saddles (\ref{correlSFF}) coincide  with  SYK correlators  on an auxiliary thermal system at inverse temperature  $\tilde{\beta}_{aux}$, computed on  the double field thermal state

\beq
| TFD \rangle  = \frac{1}{\sqrt{Z(\beta)}} \sum_{n = 1}^{2^{\frac{N}{2}}}  e^{- \frac{\beta}{2}E_n }  | E_n \rangle_{L} \otimes | E_n \rangle_{R}. \label{TFD}
\eeq

  at the auxiliary (rescaled)   inverse  temperature

\beq
\tilde{\beta}_{aux}   \equiv    \frac{\beta_{aux}}{\sqrt{1 +  \frac{\beta^2}{T^2}}}. \label{betatilde}
\eeq

The reason for that is that  saddles  are solutions of  local differential equations, and  the case of the spectral form factor and those of the auxiliary thermal system differ by boundary conditions,  by the way  the Keldysh contour for the path integral is
closed in the complex time plane. More precisely, the two points correlators of   the  spectral form factor  are complexified by the presence of the  complex  coefficients in $J_{ij}$ w.r.t.h. two points correlators over a double field thermal state.

In order to emphasize more the relation among   the conformal  saddles of the spectral form factor  (\ref{correlSFF}) and two point functions of SYK  over an auxiliary thermal state, we observe  that the former  can be written  in the following equivalent form, where
the dependence on the auxiliary system  inverse temperature   $\tilde{\beta}_{aux}$ (\ref{betatilde}) is stressed

\bea
G_{LL}^{\beta_{aux}}(t) &=&  \frac{1}{( J_{LL})^{1/4}} G^{\tilde{\beta}_{aux}}_{c}(t),  \nn \\
G_{LR}^{\beta_{aux}}(t) &=&  G_{RL}^{\beta_{aux}}(t) =    \frac{i}{ (J_{LR})^{1/4} }  \left|G^{\tilde{\beta}_{aux}}_{c}\left( t + \Delta - i \frac{\tilde{\beta}_{aux}}{2} \right)\right|, \nn \\
G_{RR}^{\beta_{aux}}(t) &=& (G_{LL}^{\beta_{aux}}(t))^* \nn \\
\label{correlSFF2}
\eea

where
\beq
G^{\beta}_{c}(t) =  \frac{1}{\sqrt{J}(4\pi)^{1/4}}
 \frac{\text{sgn}(t)}{ \left| \frac{\beta}{\pi}\sinh \left( \frac{\pi}{\beta} t  \right) \right|^{1/2} },
\eeq

is the analytic continuation to real time $\tau = it$ of the  regular SYK conformal thermal  Green function at inverse temperature $\beta$.
Let us notice that the left-right correlator $ G_{LR}^{\beta_{aux}}(t)$ in (\ref{correlSFF2})   is obtained from  the diagonal one by the shift in the real time argument    $t \rightarrow t - i \frac{\tilde{\beta}_{aux}}{2}$. This  is the  standard    prescription  that relates  a  correlator on the  purified   double field thermal state (\ref{TFD}) to  a thermal  correlator on one copy of the system.    When a  generic operator  is    moved from one copy to the other in  the purified system,  besides the imaginary   shift on the time argument, is has to be also CPT conjugated. In the SYK  case, CPT conjugation reduces to the identity  on Majorana fermions. From the bulk point of view the imaginary time shift by $\pm i \frac{\beta}{2}$ allows to map points of one side to corresponding points in the other side  in a complex time  wormhole geometry.

The meaning of the parameter  $\beta_{aux}$ is explained in \cite{Saad:2018bqo}, by considering   the spectral form factor in the  infinite temperature limit and  showing  that it  can be  approximated  by a family  of  thermal  partition functions labeled by the auxiliary inverse temperature $\beta_{aux}$.
   From a bulk  gravity  perspective, the auxiliary thermal system inverse temperature  $\beta_{aux}$ fixes the energy scale of the two holographic boundaries in $nAdS_2$.  In the $nAdS_2$  wormhole, described holographically   by the double field thermal state (\ref{TFD}), on each of the two thermal Rindler pathces, the metric and the dilaton have the following form

\beq
ds^2 =  - \sinh(\rho)^2 d\tilde{t}^2 + d \rho^2,  \qquad \Phi = \Phi_h \cosh(\rho). \label{bulkmetric}
\eeq

In the bulk  solution relevant for  the  spectral form factor  there is a  Rindler time  periodic  identification  $\tilde{t}_{R} \sim \tilde{t}_{R} + \tilde{T}$ and $\tilde{t}_L \sim \tilde{t}_L  - \tilde{T}$. This is compatible with the asymmetric analytic continuation in the spectral form factor  $\langle  Z(\beta - iT) Z(\beta + iT) \rangle_{c} $.
This Rindler time identifications  give  rise to a Lorentzian manifold with the topology of a  double cone with closed time curves \cite{Saad:2018bqo}.
$\beta_{aux}$ arises in relating Rindler time  $\tilde{T}$ to boundary time   $T$ in the following way.
By using an holographic renormalization parameter $\epsilon$ one can relate the boundary proper time $t$ to the bulk Rindler time $\tilde{t}$ from  the bulk   metric (\ref{bulkmetric}) at a large fixed   $\rho = \pm \rho_c$

\beq
d t^2 =  \epsilon d s^2 |_{\rho = \pm \rho_c }.
\eeq

This  gives the following  relation between SYK boundary time $t$ and bulk Rindler time $\tilde{t}$
\beq
t = \epsilon \frac{e^{\rho_c}}{2} \tilde{t} = \frac{\beta_{aux}}{ 2\pi} \tilde{t}.
\eeq
This relation was also checked numerically in  \cite{Saad:2018bqo}, by  computing geodesics distances from boundaries points in the  $AdS_2$ wormhole geometry and getting  agreement with (\ref{correlSFF}).
The idea is that in the large $N$ limit the bulk theory classicizes and boundary CFT two points  correlators reduce to simple functions  of the length of the geodesic connecting the pair of points on which the two point function is evaluated.
Besides  $\beta_{aux}$, there is at least a second phase space parameter, since  phase space is always even dimensional. In the case of pure  JT gravity the dimension of phase space is two, (see for example \cite{Harlow:2018tqv}  for a detailed account).
This missing phase space parameter is  the compact parameter $\Delta \in [- T, T]$ that appears in the off diagonals saddle correlator $G_{LR}^{\beta_{aux}} = G_{RL}^{\beta_{aux}}$ in  (\ref{correlSFF}) and (\ref{correlSFF2}) and it is responsible for the linear ramp behavior in the SYK spectral form factor \cite{Saad:2018bqo}.
$\Delta$ corresponds to a  relative shift between the  time coordinate origins on the right and left boundary.
It is not a  surprise that it appears only on the left-right  diagonal  correlator in (\ref{correlSFF}), since only there a relative shift on the origins of times coordinates is relevant.
The fact that $\Delta$ is responsible for the ramp goes as follows \cite{Saad:2018bqo}. It turns out that the conformal saddles (\ref{correlSFF}) have zero action. In order to compute the functional integral in the large $N$ limit,  one  has to still integrate over the phase space parameters  $\beta_{aux}$ and $\Delta$.  Integration over $\beta_{aux}$  gives just a constant  overall constant to the spectral form factor. On the other hand,  $\Delta$ integration   on  $[-T, T]$   gives a linear  $T$ factor which reproduces  the ramp.
The above discussion was at the conformal level, for the spectral form factor family of conformal  saddle points of the   critical action.
When effects of the two replica bilocal  kinetic  operator $\hat{\sigma}_{ij}(\tau, \tau')$ (\ref{kin}) are taken into account, at lowest order in the perturbative expansion an effective action for the left and right time reparametrization  soft modes $f_{L}(\tau)$, $f_{R}(\tau)$ occurs. This   effective action at the saddle point level turns out to be independent on $T$, consistently with the existence of a linear ramp. However, the value of the action in the saddle point is non zero and a  runaway direction for the functional integral representation of the spectral form factor arises, toward  large $\beta_{aux}$.
The $\beta_{aux}$  instability  can be cured in  the microcanonical ensemble \cite{Saad:2018bqo}, this is a consequence of  $\beta_{aux}$  being  actually  related to the energy of the Schwarzian modes. On the other hand, $\beta_{aux}$ can also be stabilized in the canonical ensemble,   at the price of slightly changing  the model,  by adding a small non local  coupling between  left and right soft modes $f_{L}(\tau)$ and $f_{R}(\tau)$  \cite{Saad:2018bqo}.  This is an interesting possibility that deserves further studies.  The two boundaries non local coupling term that stabilizes $\beta_{aux}$
has some formal analogy    to   non-local  double trace  interaction terms that  make a wormhole  temporarily traversable \cite{Gao:2016bin},       \cite{Maldacena:2017axo}. Analysis  related to this discussion  from the perspective  of  two boundaries  JT gravity  is found in \cite{Blommaert:2018iqz}.
 Beyond the Schwarzian saddle point approximation, the one loop determinant receives  contributions both  from the two time  reparametrization soft modes $f_{L}(\tau)$, $f_{R}(\tau)$,
and   from fluctuations of the phase space parameters $\delta \beta_{aux}(t)$, and $\delta \Delta (t)$ \cite{Saad:2018bqo}. These latter fluctuations  are controlled by an hydrodynamic  action and  computation of the one loop determinant consistently reproduces the ramp \cite{Saad:2018bqo}.

\section{A large $N$, large $T$, approximation for the spectral form factor path integral} \label{sec2}

In the first part of this section,  we obtain   a large $N$ approximation  of the  spectral form factor path integral by using steepest descent method   through the replica non-diagonal conformal saddle (\ref{correlSFF}). We then obtain a large $T$ approximation of what we got, by computing the effects of a regularization of
the kinetic operator $\hat{\sigma}_{ij}  =  \delta_{ij}\delta(t -t')\partial_{t}$.
The obtained results allow us  then to find
the  large $T$ effective action for the spectral form factor in the   time-reparametrization soft modes $f_{L}(t)$ and $f_{R}(t)$.

\vspace{.3 cm}

By performing  a     translation in field space  $\Sigma_{ij}(t, t') \rightarrow  \Sigma_{ij}(t, t') +  \hat{\sigma}_{ij}(t, t')$,
where $\hat{\sigma}_{ij}(t, t') = \delta_{ij} \delta(t - t' ) \partial_{t}$,
  the action (\ref{I}) turns into

\bea
 && \frac{I(G_{ij},\Sigma_{ij}) }{N} =  - \frac{1}{2} \log \det \left(  - \Sigma_{ij}(t, t')  \right)   \nn \\
  &+& \frac{1}{2}\sum_{i,j} \int_{0}^{T} d t d t'  \left( \Sigma_{ij}(t, t' ) G_{ij}(t, t') - \frac{J^2 J_{ij} }{4}  G_{ij}(t, t')^4 \right)
+ \frac{1}{2}\sum_{i,j} \int_{0}^{T} d t d t'  \hat{\sigma}_{ij}(t, t' ) G_{ij}(t, t'). \nn  \\ \label{Ishifted}
\eea

As it was remarked in the previous section, (see the discussion that leads to eq. (\ref{Gtransf})),  the first three terms of the above action are   invariant under the simultaneous time
reparametrization $t \rightarrow  f_{1}(t)$,  $t' \rightarrow f_{2}(t')$, where $G_{ij}(t_1, t_2)$ transforms as a primary bilocal field with weights $\Delta = \frac{1}{4}$ and    $\Sigma_{ij}(t_1, t_2)$ transforms with weights $1 - \Delta$.
The spectral form factor conformal saddle  (\ref{correlSFF}) or (\ref{correlSFF2})  break spontaneously the twofold $SL(2, \mathbb{R})_{L} \times SL(2,\mathbb{R})_{R}$ symmetry down to  the  diagonal $SL_{diag}(2,\mathbb{R})$ subgroup,  a fact that   can be checked directly by acting with Moebius transformations on the correlators in (\ref{correlSFF}) or (\ref{correlSFF2}).
   On the other hand, the source term in (\ref{Ishifted}) breaks explicitly conformal invariance. In order to study the effects of the kinetic operator $\hat{\sigma}_{ij}(t, t')$,  we shall  regularize it, by smearing   the singular Dirac delta kernel on the  account that times   shorter   than the time scale related to the   energy of  the system cannot be resolved.

In the large $N$ limit, at strong coupling,  the   path integral for the spectral form factor can be approximated by using  the steepest descent method
   through the  conformal  saddle  (\ref{correlSFF2})  $(G_{ij}^{c}, \Sigma_{ij}^{c})$

\bea
G_{ij}  &=& G_{ij}^c +     \frac{1}{J_{ij}^{1/2} |G_{ij}^c|} \delta G_{ij} \nn \\
\Sigma_{ij}  &=& \Sigma_{ij}^c +   J_{ij}^{1/2} |G_{ij}^c|  \delta \Sigma_{ij}, \nn \\ \label{fluctSFF}
  \eea

 where the  normalization in front of the fluctuations is chosen  for convenience.  %The next step is
An expansion of the action up  to quadratic order in the fluctuations gives

\bea
\frac{I}{N}  &\sim & \frac{I_{c}}{N} -  \frac{1}{12J^2 }\langle \delta \Sigma_{ij} | \tilde{K}_{ij,kl}^{c} | \delta \Sigma_{kl} \rangle + \frac{1}{2} \langle \delta \Sigma_{ij} |\delta G_{ij} \rangle   - \frac{3}{4}J^2  \langle \delta G_{ij} | \delta G_{ij} \rangle \nn \\
&&  + \frac{1}{2} \left\langle \sigma_{ij} |  G_{ij}^{c} + \frac{\delta G_{ij}}{J_{ij}^{1/2}|G_{ij}^c|} \right\rangle, \label{IGSigma1}
 \eea
where  $I_{c}$ is the action evaluated in the  conformal saddle, replica  indexes $i,j = R,L$ are summed over and  $\langle f  |  g \rangle$ denotes the following  scalar  product in  the space of bilocal functions

\beq
\langle f  |  g \rangle \equiv  \int d t d t'  f^{*}(t, t') g(t, t').
\eeq

 The four indexes  integral kernel appearing  in (\ref{IGSigma1})

\beq
\tilde{K}^{c}_{ij, kl}(t_1, t_2 ; t_3, t_4) =  3 J^2 J_{ij}^{1/2}J_{kl}^{1/2} |G_{ij}^{c}(t_{12})|  G_{ik}^{c}(t_{13})G_{jl}^{c}(t_{24})|G_{kl}^{c}(t_{34})|,   \qquad  t_{pq} = t_p - t_q   \label{Kij}
\eeq

  is a colored version  of the four point function symmetrized ladder kernel in regular SYK \cite{Maldacena:2016hyu}. The latter for generic even coupling $q \ge 4$ has the form

\beq
\tilde{K}(t_1, t_2 ; t_3, t_4) =   J^2 (q - 1) |G^{\frac{q-2}{2}}(t_{12})|  G(t_{13})G(t_{24})|G^{\frac{q-2}{2}}(t_{34})|. \label{symKSYK}
\eeq

We now introduce a convenient  alternative   indexing for the replica entries. We define   $\alpha = 1,2,3,4$  such that  $\alpha = 1 = LL$, $\alpha = 2 = RR$, $\alpha = 3 = LR$, $\alpha = 4 = RL$. The action (\ref{IGSigma1}) in this  new indexing reads

\bea
\frac{I}{N}  &=& \frac{I_{c}}{N} -  \frac{1}{12J^2 }\langle \delta \Sigma_{\alpha} | \tilde{K}_{\alpha \beta }^{c} | \delta \Sigma_{\beta} \rangle + \frac{1}{2} \langle \delta \Sigma_{\alpha} |\delta G_{\alpha} \rangle   - \frac{3}{4}J^2  \langle \delta G_{\alpha} | \delta G_{\alpha} \rangle \nn \\
&&  + \frac{1}{2} \left\langle \sigma_{\alpha} |  G_{\alpha}^{c} + \frac{\delta G_{\alpha}}{J_{\alpha}^{1/2} |G_{\alpha}^c|} \right\rangle \label{IGSigma3}.
 \eea

By  integrating    along the steepest descent direction in the   $\delta \Sigma_{\alpha}$  complex plane, passing through the saddle $\Sigma_{\alpha}^{c}$,  one  finds

  \bea
\frac{\tilde{I}}{N}  &=&  \frac{I}{N} - \frac{I_c}{N}  \nn \\
 &=&
  \frac{3}{4}J^2 \left\langle \delta G_{\alpha} | \left( (\tilde{K}_{c}^{-1})_{\alpha \beta} - \delta_{\alpha \beta }  \right) | \delta G_{\beta} \right\rangle   \nn \\
 &+& \frac{1}{2}  \left \langle \sigma_{\alpha} |   \frac{\delta G_{\alpha}}{J_{\alpha}^{1/2} |G_{\alpha}^c|} \right \rangle +  \frac{1}{2} \langle \sigma_{\alpha}  | G_{\alpha}^{c} \rangle. \nn \\ \label{IG}
\eea

By then integrating along the steepest descent contour in the   $\delta G_{\alpha}$  complex plane through the saddle $G_{\alpha}^{c}$, one finds the following expression for the action  of the spectral form factor in the large $N$ approximation

\beq
\frac{\tilde{I}}{N} =  \frac{1}{12 J^2} \langle s_{\alpha} |  \left( \delta_{\alpha \gamma} -  \tilde{K}_{\alpha \gamma }^{c}   \right)^{-1} \tilde{K}^{c}_{\gamma \beta}  | s_{\beta}\rangle  + \frac{J_{\alpha}^{1/2}}{2}\langle s_{\alpha}  | |G_{\alpha}^{c}|^2 \rangle,  \label{Isource}
\eeq

 where   $s_{\alpha}(t, t')$ is the rescaled   source

 \beq
  s_{\alpha}(t, t') \equiv \frac{\sigma_{\alpha}(t, t')}{J_{\alpha}^{1/2} |G_{\alpha}^{c}(t, t')|}. \label{s}
 \eeq

So far, by steepest descent  we  got  to the following large $N$ approximation for the spectral form factor

\beq
Z(\beta + iT)Z(\beta - iT) \sim  e^{ \frac{N}{12 J^2} \langle s_{\alpha} |  \left( \delta_{\alpha \gamma} -  \tilde{K}_{\alpha \gamma }^{c}   \right)^{-1} \tilde{K}^{c}_{\gamma \beta}  | s_{\beta}\rangle  + \frac{N}{2}\langle s_{\alpha}  | J_{\alpha}^{1/2}|G_{\alpha}^{c}|^2 \rangle}. \label{Z2approx}
\eeq

In principle,  (\ref{Z2approx}) allows to compute   quantum corrected forms of various correlators by functional deriving w.r.t. regularized  source $\sigma_{\alpha}$. In particular, the two  point function is given by

\bea
\langle G_{\alpha} \rangle &=& \frac{1}{N} \frac{\delta |Z|^2}{\delta \sigma_{\alpha} } =  \frac{1}{N}\frac{1}{J_{\alpha}^{1/2}|G_{\alpha}^{c}|} \frac{\delta |Z|^2}{\delta s_{\alpha} } \nn \\
& \sim & \left[ \frac{1}{2} G_{\alpha}^{c}  +  \frac{1}{12 J^2 J_{\alpha}^{1/2}|G_{\alpha}^{c}|   }  \left( \delta_{\alpha \gamma} -  \tilde{K}_{\alpha \gamma }^{c}   \right)^{-1} \tilde{K}^{c}_{\gamma \beta}  | s_{\beta}\rangle \right] e^{N I}. \label{Ga}
\eea

However, our analysis needs to be developed further, since  a suitable  regularization  for the source (\ref{s}) appearing  in  (\ref{Z2approx}) and (\ref{Ga}) are still to be discussed.  We are going to do it  now and, in particular,  we will  find   explicit results for  the large time  $T$  regime.
We  consider the eigenvalues problem  for  the spectral form factor four points function conformal ladder  Kernel

\beq
 \tilde{K}^{c}_{\alpha \beta}(t_1, t_2 ; t,t')\psi_{\beta}^{h}(t,t') =  \tilde{k}_{SFF}(h)\psi_{\alpha}^{h}(t_1 ,t_2).  \label{eighenfK}
\eeq

Where, $h$ labels  the eigenvalues of the Carimir $\mathcal{C}_{diag}$ of $SL_{diag}(2)$, the diagonal subgroup of  $SL_{L}(2) \times S_{R}(2)$ which  is the symmetry group  that survives  the  spontaneous breaking induced   by the conformal  replica non-diagonal saddles. The   $h$ dependence on the eigenvalues $\tilde{k}_{SFF}(h)$ is due to the fact that  $\tilde{K}^{c}_{\alpha \beta}(t_1, t_2 ; t,t')$ commutes with  $\mathcal{C}_{diag}$.

We  assume the following expansion  for the source,  in terms of the ladder kernel  eigenfunctions in (\ref{eighenfK})

\beq
s_{\alpha}(t,t') = \sum_{h} C_{h} \psi_{\alpha}^{h}(t,t') U(\log \zeta_{\alpha}(|t - t'|)). \label{sou}
\eeq

In the regulirized rescaled kinetic operator (\ref{sou}),  $U(\xi)$ is a smooth smearing  function of  $\xi_{\alpha} = \log( \zeta_{\alpha}(| t - t'|)) $   that regularizes the singular Dirac kernel $\delta(t - t')$  in the two replica kinetic operator $\hat{\sigma}_{ij} =  \delta_{ij}\delta(t - t')\partial_{t}$.
$U$ cuts off  the  short times intervals  region below the scale related to the SYK energy scale $1/J$.  It works as follows, $U(\xi)$ is demanded to   vanish  for $\xi < 0$, and for $|t - t'| > 2T$.   $\zeta_{\alpha}(|t - t'|)$ is defined to be the function appearing  in the  denominator of the $\alpha$ component of the weight $h$ eigenfunction of the colored kernel

\beq
\psi_{\alpha}^{h}(t,t') \propto  \frac{1}{\zeta_{\alpha}(|t - t'|)^{h}}.  \label{struct}
\eeq

 Since  $\psi_{\alpha}^{h}(t,t')$ is  also an eigenfunction of the $SL_{diag}(2)$
Casimir $\mathcal{C}_{diag}$,  $\zeta_{\alpha}(|t - t'|)$ is typically an hyperbolic sine  or cosine.
One has   $0 < \zeta_{\alpha}(|t - t'|) \le 1$ for  $|t - t'| <  \frac{1}{J}$ , which makes $\xi_{\alpha} < 0$ because of the logarithm and the smearing function $U$  vanishes correspondingly. This implements the impossibility to resolve time intervals below $1/J$.
On the other hand, for large enough  $|t - t'|$,    $\zeta_{\alpha}(|t - t'|)$  grows exponentially in the separation time.
This makes  the support of the smearing function $U(\log( \zeta_{\alpha}(| t - t'|)))$ to be of order $O(T)$.

We also demand the following normalization for the smearing function

\beq
\int d \xi \,  U(\xi)  = 1.
\eeq

Let us  write  the smearing function as a Fourier integral

\beq
U(\xi) = \int \frac{d \eta}{ 2 \pi} e^{- i \xi \eta} \tilde{U}(\eta). \label{fourier}
\eeq

Since $ U(\xi)$ has support over  $\Delta \xi \sim O(T) $,
  its   Fourier transform  $\tilde{U}(\eta)$
has support   over   $\Delta \eta  \sim  O\left( \frac{1}{T}\right)$. In the  large $T$ limit, the Fourier transform $\tilde{U}(\eta)$ becomes very narrow.

By inserting   (\ref{sou})  and (\ref{fourier})    into   (\ref{Ga}), one finds

\beq
\langle G_{\alpha}(t,t') \rangle  \sim  \left[ \frac{1}{2} G_{\alpha}^{c}  +  \frac{1}{12 J^2 J_{\alpha}^{1/2}|G_{\alpha}^{c}|   } \sum_{h} c_{h} \int \frac{d \eta}{ 2\pi}  \frac{\tilde{k}_{SFF}(h + i\eta )}{1 - \tilde{k}_{SFF}(h + i \eta)} \psi_{\alpha}^{h + i \eta }(t,t') \tilde U(\eta)    \right] e^{N I}, \label{Ga2}
\eeq

where we used  both
\beq
e^{- i \eta \xi}  =  \frac{1}{ \left(\zeta_{\alpha}(|t - t'|)^{h} \right)^{i\eta}},
\eeq

and eq. (\ref{struct}).

 Since  $\tilde{U}(\eta)$  has support   $\Delta  \eta =  O\left( \frac{1}{T} \right)$,
 in the  large $T$ limit,  one can  expand   $\tilde{k}_{SFF}(h + i\eta)$   up to the first order in eta
$\tilde{k}_{SFF}(h + i\eta) = \tilde{k}_{SFF}(h) + i \eta \, \tilde{k}_{SFF}'(h)$. Then,   by computing the $\eta$ integrals in  (\ref{Ga2}) by   residues theorem,  one finds

\beq
\langle G_{\alpha}(t,t') \rangle  \sim  \left[ \frac{1}{2} G_{\alpha}^{c}  +  \frac{i}{12 J^2 J_{\alpha}^{1/2}|G_{\alpha}^{c}|   } \sum_{h} c_{h}   \psi_{\alpha}^{h + i \eta_h }(t,t') \tilde U_{h}(\eta_h )    \right] e^{N I},  \label{Ga3}
\eeq

with

\beq
   \eta_h = i \frac{\tilde{k}_{SFF}(h) - 1}{\tilde{k}_{SFF}'(h)}. \label{etah}
\eeq

Since the support of $\tilde{U}(\eta)$ is $\Delta \eta  \sim  O\left(\frac{1}{T}\right)$,  for  large enough $T$,
the only non vanishing term in (\ref{Ga3}) is the  $\eta_{h} = 0$  one, which by  (\ref{etah})   corresponds to  the $\tilde{k}_{SFF}(h_{*}) = 1$ eigenvalue.

To summarize, we have shown that in the large $N$, strongly coupled regime, the SYK  spectral form factor at large $T$
is approximated by (\ref{Z2approx}) with a regularized  source  invariant under the four point ladder kernel.
In particular, the large $T$ limit of the  two point function is given by
\beq
\langle G_{\alpha}(t,t') \rangle  \sim  \left[ \frac{1}{2} G_{\alpha}^{c}  +  \frac{i}{12 J^2 J_{\alpha}^{1/2}|G_{\alpha}^{c}|   }    \psi_{\alpha}^{h_{*}}(t,t')    \right] e^{N I},
\eeq

where  $\psi_{\alpha}^{h_{*}}(t_1 ,t_2)$

\beq
 \tilde{K}^{c}_{\alpha \beta}(t_1, t_2 ; t,t')\psi_{\beta}^{h_{*}}(t,t') =  \psi_{\alpha}^{h_{*}}(t_1 ,t_2),
\eeq

is the  invariant function  under the ladder colored  conformal  kernel.
It follows  that  at large $T$, the regularized   source
has the form

\beq
s_{\alpha}(t,t') = \mathcal{P} \left( \psi_{\alpha}^{h_{*}}(t,t') \right) U\left(\log \left( J \frac{\tilde{\beta}_{aux}}{\pi}\sinh \frac{\pi}{\tilde{\beta}_{aux} } |t - t'| \right)\right), 
\eeq

where $\mathcal{P}$ is the projector over the two  diagonal directions $\alpha = 1 = LL$ and $\alpha = 2 = RR$ in replica indexes.
We impose this projection,   since  the source is a regularized version  of the \emph{replica diagonal} SYK kinetic operator   $\delta(t_{i} - t_{i}')\partial_{t_{i}}\delta_{ij}  $, $i= L,R$.

Since $s_{1}(t,t') = s_{LL}(t,t')$ is expected to have the same form as  $s_{2}(t,t') = s_{RR}(t,t')$,
it follows that the source that dominates the large $T$ dynamics needs to be invariant
under  the operator $\mathcal{P}^{-1} \tilde{K}^{c}_{\alpha \beta} \mathcal{P} = \tilde{K}^{c}_{11}$.

By using  (\ref{correlSFF2}) in (\ref{Kij})  one finds that   $\tilde{K}^{c}_{11} =\tilde{K}^{c}_{LLLL} = \tilde{K}^{c}_{22} = \tilde{K}^{c}_{RRRR} =  \tilde{K}_{c}^{\tilde{\beta}_{aux}}$,  where  $\tilde{K}_{c}^{\tilde{\beta}_{aux}}$ is the regular SYK,  $q = 4$,  four point function symmetrized ladder kernel (\ref{symKSYK})  at inverse temperature $\tilde{\beta}_{aux}$

\bea
\tilde{K}^{c}_{11}(t_1, t_2 ; t,t') &=& 3J^2 J_{LL} |G_{LL}(t_1,t_2)| G_{LL}(t_1,t) G_{LL}(t_2, t') |G_{LL}(t,t')| \nn \\
&=&   3J^{2}|G_{c}^{\tilde{\beta}_{aux}}(t_1,t_2)| G_{c}^{\tilde{\beta}_{aux}}  (t_1,  t) G_{c}^{\tilde{\beta}_{aux}}  (t_2, t') |G_{c}^{\tilde{\beta}_{aux}}(t,t')| \nn \\
 &=& \tilde{K}_{c}^{\tilde{\beta}_{aux}}(t_1, t_2; t, t')
\eea

with

\beq
G_{c}^{\beta}(t,t') = \frac{1}{(4\pi)^{1/4}\sqrt{J}} \frac{sgn(t-t')}{\left| \frac{\beta}{\pi}  \sinh (\frac{\pi}{\beta}(t - t')) \right|^{1/2}}.
\eeq

Therefore, the regularized source  that  dominates  the spectral form factor  at large $T$ can be  obtained  by looking
for  invariant function under the four point ladder conformal  kernel in regular SYK

\beq
\tilde{K}_{c}^{\beta}(t_1, t_2; t,t') \psi_{h_{*}}(t,t') =   \psi_{h_{*}}(t,t').
 \eeq

The spectrum of eigenvalues and eigenfunctions of
 of $\tilde{K}_{c}^{\beta}$ are known

 \beq
\tilde{K}_{c}^{\beta}(t_1, t_2; t,t') \psi_{h}(t,t') = \tilde{k}_{c}(h)  \psi_{h}(t,t'), \label{eigheq}
\eeq

in the following we recall some of their properties.
 Let us notice that the  eigenvalues  are actually   independent on $\beta$.
One can check by using the explicit for of ladder kernel that  by  a time rescaling
 one can change the  value of $\beta$ and  leaves   the form of the   eigenvalues  equation (\ref{eigheq}) invariant.
 In fact, a time rescaling is just an  $SL(2)$ transformation.
It  is therefore possible  to    consider (\ref{eigheq}) either on   the time line at zero temperature, or at fine temperature at imaginary time
of at fine temperature in real time.
 Eigenfunctions at finite $\beta$, are then found  by applying  the usual  conformal transformation $t = \tan\left( \frac{\pi}{\beta} \tau \right)$ on those defined on the real line.
On the other hand,   real time eigenfunctions at finite temperature  are obtained by  analytic continuation of the imaginary time ones.

Since the $SL(2)$      Casimir  operator  $\mathcal{C}$  commutes with the four points ladder conformal  kernel  $ \tilde{K}_{c}$,  $\mathcal{C}$ eigenfunctions

\beq
\mathcal{C}(t_1, t_2 ; t, t' ) \psi_{h} (t,t') = h(h - 1) \psi_{h}(t_1 ,t_2 ).
\eeq
are also $ \tilde{K}_{c}$ eigenfunctions.
$\mathcal{C}$  has a spectrum with   both a  continuum and a discrete component  \cite{Maldacena:2016hyu}. The continuum component  is given by the points on the   critical strip  $h_s  =  \frac{1}{2} + is$, $s \in \mathbb{R}$, while   the discrete component is given by  $h_{n}  =  2n$,  $n \in \mathbb{N}_{> 0}$.
      The  Casimir $\mathcal{C}$  eigenfunctions  have the following form \cite{Maldacena:2016hyu}

\beq
\psi_{h}(t_0, t_1, t_2) = \int d t_0  g_{h}(t_0) \frac{\text{sgn}(t_{12})}{ |t_{01}|^{h} |t_{02}|^{h} | t_{12}|^{1- h}}. \label{eighenf2}
\eeq

Notice that in the  $t_0$ large limit

\beq
\psi_{h}(t_0, t_1, t_2) \sim  \frac{sgn(t_{12})}{| t_{12}|^{1- h}},   \qquad  t_0 \rightarrow \infty
\eeq

In the eighenvalue equation
  \beq
\tilde{K}_{c}(t_1 , t_2 ; t, t' ) \psi_{h}(t_0,   t , t' ) =  \tilde{k}_c (h) \, \psi_{h}(t_0,  t_1 , t_2 ),
 \eeq

by $SL(2)$ symmetry, one can fix $t_0 = \infty$, $t_1 = 1$ and $t_2 = 0$  and
 compute    explicitly   $\tilde{k}_{c}(h)$ by  the following  double integral

 \beq
 \tilde{k}_{c}(h) = \tilde{K}_{c}(1 , 0 ; t, t' ) \psi_{h}(\infty,   t , t' ).
 \eeq

For the  four order coupling   $q=4$,  the result is

\beq
\tilde{k}_{c}(h) =  - \frac{3}{2} \frac{\tan \left(\frac{\pi}{2} \left( h - \frac{1}{2} \right) \right)  }{ h - \frac{1}{2}}. \label{kc2}
\eeq

The discrete component  of the spectrum  $Spec(\mathcal{C})$ thus is   given

\beq
 \tilde{k}_c (2n) = \frac{3}{4n - 1}   \qquad  n =1,2,\dots. \label{discrete}
\eeq

We are looking for   invariant functions under the action of $\tilde{K}_c$.  From the above  results on the spectrum, we have
that   $\tilde{k}_{c}(2) = 1$.

On the other hand, for  the continuum component of the spectrum one finds

\beq
k_{c}(h_{s}) = i \frac{3}{2} \tanh \left( \frac{\pi}{2} s \right),    \qquad  s \in \mathbb{R} \label{continuum}
\eeq

 does  not provide any further invariant function, besides  the one that we found from the discrete component of the Casimir spectrum.

To summarize, we have found   that  the contributions to the   source that dominates   the large  $T$ regime for  the spectral form factor has the following form

\beq
s_{LL}(t, t') = s_{RR}(t, t')  =  a_0  \frac{\text{sgn}(t - t')}{\left[ J\frac{\tilde{\beta}_{aux}}{\pi} \sinh \left( \frac{\pi}{\tilde{\beta}_{aux}} |t - t' | \right)\right]^{2}} U\left(2  \log \left( J \frac{\tilde{\beta}_{aux}}{\pi}\sinh\left( \frac{\pi}{\tilde{\beta}_{aux} } |t - t'|\right) \right)\right), \label{sexpression2}
\eeq

where  $a_0$ is a coefficient to be fitted numerically,

\subsection{Effective action for  the time reparametrization soft modes}

 In order to find the large $T$ spectral form factor  effective action     in the time reparametrization soft modes  $f_{L}(t)$, $f_{R}(t)$,   we consider
 the conformal symmetry breaking source term  in (\ref{Ishifted})

\beq
\frac{I_{s}}{N} = \int_{0}^{T} d t d t' \left( G_{LL}(t, t') \sigma_{LL}(t, t')  +   G_{RR}(t, t') \sigma_{RR}(t,t')\right)
  \label{source2}
\eeq

and take the diagonal conformal symmetry  breaking source,   obtained in the previous section in (\ref{sexpression2})

\beq
\sigma_{LL}(t, t') =   \frac{a_0}{J_{LL}^{1/4}}  \frac{\text{sgn}(t - t')}{\left[ J \frac{\tilde{\beta}_{aux}}{\pi} \sinh \left( \frac{\pi}{\tilde{\beta}_{aux}} |t - t' | \right)\right]^{5/2}} U\left(\log  \left( J \frac{\tilde{\beta}_{aux}}{\pi} \sinh\left( \frac{\pi}{\tilde{\beta}_{aux} } |t - t'| \right)\right)\right). \label{sexpression3}
\eeq

On the other hand,  $\sigma_{RR}(t, t')$ has  the same form (\ref{sexpression3}) but with  $J_{RR}$  replacing  $J_{LL}$.

The components of the  Green function by a time reparametrization  $t_{i} \rightarrow f_{i}(t_i)$ transform as

\beq
G_{ii}(t, t') =  \frac{1}{(J_{ii})^{1/4}}G_{\beta = \infty}(f_{i}(t), f_{i}(t')) f_{i}'(t)^{\Delta} f_{i}'(t')^{\Delta},  \qquad  \Delta = \frac{1}{4}  \qquad i=L,R, \label{soft2}
\eeq

where
\beq
G_{\beta = \infty}(t, t' ) =  \frac{1}{(4\pi)^{1/4}\sqrt{J}} \frac{\text{sgn}(t - t')}{ |t - t'|^{1/2} }.
\eeq

The breaking of conformal invariance in the SYK model is due to a uv short-time effect, as the conformal forms for the correlators describe accurately the low energy ir dynamics.     In order to find a local   effective local action,
we go from  $(t, t')$  to $(t_{+}, t_{-})$, slow average time $t_{+} = \frac{t + t'}{2}$, and  $t_{-} = t - t'$.
We expand the fields up to lowest order in $t_{-}$ and then we  integrate out $t_{-}$.
This gives a   local effective action in  $t_{+}$ in  the two  time reparametrization soft modes $f_{i}(t^{+}_{i})$, $i = L,R$.

A Taylor expansion in $t_{-}$ gives
 \beq
G_{ii}(t_{+}, t_{-}) =  \frac{1}{(J_{ii})^{1/4}} G_{\beta = \infty}(i t_{-}) \left( 1 + \frac{\Delta}{6} \text{Sch} (f_{i}(t_{+}), t_{+}) t_{-}^2  + O\left( t_{-}^4  \right)  \right). \label{expansG2}
 \eeq

On the other hand,   the diagonal components   (\ref{correlSFF2})  of the  conformal spectral form factor  Green functions,
obtained from (\ref{soft2}) for $f_{L}(t) = \tanh\left( \frac{\pi}{\tilde{\beta}_{aux}} t \right)$ and  $f_{R}(t) = - \frac{1}{ \tanh\left( \frac{\pi}{\tilde{\beta}_{aux}} t \right)}$
 have the explicit form

\beq
G^{c}_{ii}(t, t') =   \frac{1}{ (J_{ii})^{1/4} (4 \pi)^{1/4} \sqrt{J}} \frac{\text{sgn}(t - t')}{\left[ \frac{\tilde{\beta}_{aux}}{\pi} \sinh \left( \frac{\pi}{\tilde{\beta}_{aux}} |t - t' | \right)\right]^{1/2}}, \qquad i=L,R. \label{thermalcG2}
\eeq

By inserting in eq.  (\ref{source2})   the short time $t_{-}$ expansions  (\ref{expansG2}),  the  rescaled  source along the enhanced direction  (\ref{sexpression3})  and   the  conformal  Green function (\ref{thermalcG2})  at inverse temperature $\tilde{\beta}_{aux}$ (\ref{betatilde})   one finds

\bea
- &\frac{I_{s}}{N}& = \frac{a_{0} (4\pi)^{-1/4}}{24 \sqrt{J J_{LL}}} \int_{0}^{T} d t_{+} \text{Sch} (f_{L}(t_{+}), t_{+})
 \int_{- T}^{T} d t_{-} G_{\beta = \infty} ( i t_{-})t_{-}^{2} \frac{\text{sgn}(t_{-})}{ \left|J \frac{\tilde{\beta}_{aux}}{\pi} \sinh\left( \frac{\pi}{\tilde{\beta}_{aux}} t_{-} \right) \right|^{2 + 1/2}}U\left(\xi(|t_{-}|)  \right) \nn \\
 &+&   L  \leftrightarrow R  \nn \\
 &\sim& \frac{\alpha_{S}}{\mathcal{J}} \left( \frac{1}{\sqrt{J_{LL}}}  \int_{0}^{T} d t_{+}    \text{Sch} (f_{L}(t_{+}), t_{+})  +   \frac{1}{\sqrt{J_{RR}}}  \int_{0}^{T} d t_{+}    \text{Sch} (f_{R}(t_{+}), t_{+}) \right)        \int_{-T}^{T}   \frac{ d t_{-}}{ J|t_{-}| }U\left(\log\left(J|t_{-}| \right) \right) \nn \\
&= &  \frac{\alpha_{S}}{\mathcal{J} } \left( \frac{1}{\sqrt{J_{LL}}}  \int_{0}^{T} d t_{+}    \text{Sch} (f_{L}(t_{+}), t_{+})  +   \frac{1}{\sqrt{J_{RR}}}  \int_{0}^{T} d t_{+}    \text{Sch} (f_{R}(t_{+}), t_{+}) \right)  \nn \\
&= &  \frac{\alpha_{S}}{\mathcal{J}\left(\frac{\beta}{T} + i\right)}  \int_{0}^{T} d t  \, \,   \text{Sch} (f_{L}(t), t)  + \frac{\alpha_{S}}{\mathcal{J}\left(\frac{\beta}{T} - i\right)}    \int_{0}^{T} d t  \, \,  \text{Sch} (f_{R}(t), t).  \nn \\
\label{result}
\eea

In the above expression $\alpha_{S}$ is  found by numerical fitting.

\section{Conclusions}

In this work we derived  a  large $N$, late times $T$,  approximation of the path integral representation of the SYK spectral form factor.
By using this  result,  we  obtained   the effective action in the two  time reparametrization soft modes  for the  SYK  spectral form factor, in the large time  $T$ regime.
Our work puts on a stronger ground  \cite{Saad:2018bqo}, where the form of the two Schwarzian effective  action was heuristically guessed by an indirect argument, supported only by  numerical evidences  \cite{DS}.

\newpage

\vspace{.3 cm}
{\large{\bf Acknowledgments}}

\vspace{.3 cm}
The Author thanks Douglas Stanford for correspondence,
 Sergio Caracciolo and Mauro Pastore for collaboration at the early stages of this project.

\vspace{.3 cm}

\vspace{.3 cm}

\section{Appendix}

\subsection{Review of the analysis of responses  to a regularized source in   standard SYK}

In this section we review a  method discussed in  \cite{Kitaev:2017awl}, for obtaining the   soft mode  Schwarzian effective  action in regular SYK.
The  SYK action in terms of  collective fields,  after a   translation in field space $\Sigma(\tau, \tau') \rightarrow \Sigma(\tau, \tau') + \hat{\sigma}(\tau, \tau')$,  with   $\hat{\sigma}(\tau, \tau')  = \delta(\tau - \tau') \partial_{\tau}$   can be  casted in the following form

\bea
 && \frac{I(G,\Sigma ) }{N} =  - \frac{1}{2} \log \det \left(  - \Sigma(\tau, \tau')  \right)   \nn \\
  &+& \frac{1}{2} \int d \tau d \tau'  \left( \Sigma(\tau, \tau' ) G(\tau, \tau') - \frac{J^2  }{4}  G(\tau, \tau')^4 \right)
+ \frac{1}{2} \int d \tau d \tau'  \hat{\sigma}(\tau, \tau' ) G(\tau, \tau'). \nn  \\ \label{IshiftedReg}
\eea

Notice that   the first three terms  in (\ref{IshiftedReg}) are conformal invariant, while  the last source term in $\hat{\sigma}(\tau, \tau')$ breaks explicitly the conformal symmetry.  In the $\beta J >> 1$ regime the  last term  can be  treated  as a perturbation  over  the SYK conformal saddle.  In the  SYK model,   times shorter then $1 / J$ cannot be resolved, therefore  in the   kinetic operator $\hat{\sigma}(\tau, \tau')$ the  Dirac delta  kernel  $\delta' (\tau - \tau')$  needs to be replaced by a regularized  smoothed kernel  that vanishes for $ | \tau - \tau' | < \frac{1}{J}$

\beq
\sigma(\tau, \tau' ) = \sum_{h} a_{h} \frac{ \text{sgn}(\tau - \tau')}{| J(\tau - \tau') |^{h}} U \left( \log\left( J | \tau - \tau' | \right) \right), \label{sourceexp}
\eeq
 where the  $h = 1$ term has the  effect as the time derivative operator $\partial_{\tau}$, but,
a priori  various scaling weights $h$ might be relevant,  due to   non linear effects in the   responses to the  perturbation.
In (\ref{sourceexp}) the smearing  function $U$, replaces the singular kernel $\delta(t -t')$ of the kinetic SYK operator. $U(\xi)$  has a smooth  compact support  and  vanishes for  $\xi = \log\left( J | \tau - \tau' | \right) \le 0$. This  implements the uv cutoff for time resolutions less then $1 / J$. In the finite temperature case, the infrared cutoff  is given by the inverse temperature $\beta$, and the smearing function   $U$ needs  to have period $\beta$. 

 Moreover, the following   normalization condition   is imposed

  \beq
  \int_{0}^{\beta} d \xi \, U (\xi) = 1.
  \eeq

In order to evaluate the functional integral in the large $N$ limit we use the  steepest descent method.  The action is expanded up to quadratic order around the conformal saddle    $(G_c(\tau_1, \tau_2)   , \Sigma_c (\tau_1, \tau_2))$

\bea
G  &=& G_c + |G_{c}|^{- 1} \delta G \nn \\
\Sigma  &=& \Sigma_c +  |G_c | \delta \Sigma, \nn \\ \label{fluct}
  \eea

Let us notice  that (\ref{fluct})  preserve the measure in the functional integral.
By keeping terms up to quadratic order in the fluctuations one gets from  (\ref{IshiftedReg})

\bea
I  &=& I_{c} -  \frac{1}{12J^2 }\langle \delta \Sigma | \tilde{K}_{c} | \delta \Sigma  \rangle + \frac{1}{2} \langle \delta \Sigma |\delta G \rangle   - \frac{3}{4}J^2  \langle \delta G  | \delta G \rangle \nn \\
&&  + \frac{1}{2} \left \langle \sigma |  G_{c} + \frac{\delta G}{|G_c |} \right  \rangle, \label{IGS}
 \eea
where  $I_{c}$ is the action evaluated in the  conformal saddles and
$\tilde{K}_c$ is the symmetrized four point function ladder kernel at the conformal point
\beq
\tilde{K}_{c}(\tau_1, \tau_2 ; \tau_3, \tau_4) =   J^2 (q - 1) |G_{c}^{\frac{q-2}{2}}(\tau_{12})|  G_{c}(\tau_{13})G_{c}(\tau_{24})|G_{c}^{\frac{q-2}{2}}(\tau_{34})|, \label{kernel}
\eeq
for even integer $q \ge 4$. Here, we are focusing on the $q = 4$ case.

 $\langle f |   g \rangle$ denotes the following  scalar product on the space of antisymmetric bilocal functions

\beq
\langle f |   g \rangle  \equiv  \int_{0}^{\beta} d \tau d \tau'  f^{*}(\tau, \tau') g(\tau, \tau').
\eeq

 By integrating  $\delta \Sigma$ on  a contour in   the complex plane  along  the steepest descent direction passing through the saddle point,   one finds
  \beq
\tilde{I}  =  I - I_c =
  \frac{3}{4}J^2 \left\langle \delta G | \left(\tilde{K}_{c}^{-1} - 1  \right) | \delta G \right\rangle
 + \frac{1}{2}  \left \langle \sigma |   \frac{\delta G }{ |G_c | } \right \rangle +  \frac{1}{2} \langle \sigma   | G_{c} \rangle.
\eeq

By then integrating  $\delta G$ along the steepest descent contour in complex plane passing through the saddle point $G_c$,  one finds

 \beq
\tilde{I} =  \frac{1}{12J^2} \langle s | \tilde{K}_{c} \left( 1 -   \tilde{K}_{c}  \right)^{-1} | s \rangle  + \frac{1}{2}\langle s  | G_{c}^2 \rangle, \label{Isourcestand}
\eeq

with

\beq
 s(\tau_1, \tau_2)  =  \frac{\sigma(\tau_1 , \tau_2 )}{ |G_{c}(\tau_1, \tau_2 )|}. \label{sstandard}
\eeq

Eq. (\ref{Isourcestand}) shows  that  a component in functions space  for the   rescaled source  $s(\tau, \tau')$  along the zero mode direction

\beq
\left( 1 -   \tilde{K}_{c}  \right) \psi_{1}(\tau, \tau' ) = 0
\eeq

gives   an enhanced effect on $\tilde{I}$. Although naively it seems there is a divergence, indeed no divergences occur after
having regularized  the SYK kinetic source  by(\ref{sourceexp})

Eq. (\ref{Isourcestand})  leads to study  the $\tilde{K}_c$ spectrum

\beq
 \tilde{K}_c \psi_{k_c }(\tau, \tau') = \tilde{k}_c \psi_{k_c}(\tau, \tau' ).
\eeq

The $SL(2,\mathbb{R})$  Casimir $\mathcal{C}$ commutes with $\tilde{K}_c$, therefore    by diagonalizing $\mathcal{C}$ one diagonalizes also $\tilde{K}_c$.
On the other hand, the $SL(2)$ Casmir eigenvalues equation

\beq
\mathcal{C} \psi_{h}(\tau, \tau') = h(h - 1) \psi_{h}(\tau, \tau') \label{casim}
\eeq

gives rise to a spectrum with both a  discrete and a continuum components. The  discrete component is given by  $h_n = 2n$ with $n \in \mathbb{N}_{> 0}$, while the   continuum one  is given by points on  the critical strip   $h_s =  \frac{1}{2} + is$, $s \in  \mathbb{R}$.  The Casimir $\mathcal{C}$  equation (\ref{casim}) fixes also the form for the    eigenfunctions \cite{Maldacena:2016hyu}

\beq
\psi_{h}(\tau_0 , \tau_1, \tau_2) = \int d \tau_0  g_{h}(\tau_0) \frac{\text{sgn}(\tau_{12})}{ |\tau_{01}|^{h} |\tau_{02}|^{h} | \tau_{12}|^{1- h}}. \label{eighenf}
\eeq

By inserting   (\ref{eighenf}) in the  $\tilde{K}_c$ eigenvalues equation

\beq
\tilde{K}_c \psi_{h }(\tau_0, \tau_1, \tau_2) =  \tilde{k}_{c}(h)\psi_{h }(\tau_0,  \tau_1, \tau_2),
\eeq

 leads to  the explicit form for  the function $\tilde{k}_{c}(h)$.
In the SYK model  with  order  $q=4$  coupling,  one has

\beq
\tilde{k}_{c}(h) =  - \frac{3}{2} \frac{\tan \left(\frac{\pi}{2} \left( h - \frac{1}{2} \right) \right)  }{ h - \frac{1}{2}}, \label{kc}
\eeq

In particular   $\tilde{k}_{c}(2) = 1$. This  means that there is an  enhancement for $\tilde{I}$ (\ref{Isourcestand}) whenever the  perturbation  source $s(\tau_1, \tau_2)$  has a component along the $\psi_{h = 2}(\tau_1 , \tau_2 )$
direction in functional space.

Let us also  observe that  the conformal two point function $G_{c}(\tau_1, \tau_2)$, by a time reparametrization $\tau \rightarrow f(\tau)$ got a variation $\delta_{c} G_{c}(\tau_1, \tau_2)$ precisely along the enhancing direction

\beq
( 1 -   \tilde{K}_{c} )\delta_{c} G_{c} =  0.  \label{zeromodeG}
\eeq

In fact, a  soft mode transformation on $G(\tau, \tau')$ gives

\beq
G(\tau, \tau') =  G_{\beta = \infty}(f(\tau), f(\tau')) f'(\tau)^{\Delta} f'(\tau')^{\Delta},  \qquad  \Delta = \frac{1}{4}   \label{soft1}
\eeq

where
\beq
G_{\beta = \infty}(\tau, \tau' ) =  \frac{1}{\sqrt{J}} \frac{\text{sgn}(\tau - \tau')}{ |\tau - \tau'|^{1/2} }.
\eeq

The conformal finite temperature Green function $G_{c}(\tau, \tau')$ is obtained by (\ref{soft1}) for $f(\tau) = \tan \left(\frac{\pi}{\beta} \tau \right)$,
its explicit form is given by

\beq
G_{c}(\tau, \tau') = \frac{1}{\sqrt{J}} \frac{\text{sgn}(\tau - \tau')}{\left[ \frac{\beta}{\pi} \sin \left( \frac{\pi}{\beta} |\tau - \tau' | \right)\right]^{1/2}}.
\eeq

An infinitesimal conformal transformation  $\delta_{c}G_{c}$   is given by the transformation law (\ref{soft1}) applied on $G_{c}(\tau, \tau')$ for a periodic soft mode $f: S^{1}_{\beta} \rightarrow S^{1}_{\beta}$  closed to the identity map   $f(\tau) = \tau + \epsilon(\tau)$.
In order to  see that eq. (\ref{zeromodeG}) holds,  let us consider the two conformal Schwinger-Dyson equations that are obtained as saddle point equations  from  the
action  (\ref{Isourcestand}) without the source term

\bea
G_{c}(\tau, \tau'') *  \Sigma_{c}(\tau'', \tau') &= & \delta(\tau - \tau'),  \nn \\
\Sigma(\tau, \tau' ) &=& J^{2} G(\tau, \tau' )^{3}, \nn \\ \label{SDyson}
\eea

where $*$ denotes the convolution product.

 These equations are derived from a  time reparametrization invariant action
and are therefore themselves  time reparametrization invariant. Therefore a variation by  $\delta_{c}$
on both  the Schwinger-Dyson equations (\ref{SDyson}) gives zero.
In particular a variation  of the  first  of the two  Schwinger Dyson equations
gives
\beq
 \delta_{c} ( G_{c} * \sigma_{c} ) =   \delta_{c}G_{c} * \Sigma_{c} + G_{c} * \delta_{c} \Sigma_{c} = 0. \label{SDvarreg}
\eeq

We then  convolute the above equation by $G_c$, which  is the inverse of $\Sigma_c$ w.r.t. the convolution product $*$,   $G_{c} = \left(\Sigma_{c}\right)^{-1}$ and get

\beq
\delta_{c}G_{c} + G_{c} * \delta_{c} \Sigma_{c} *  G_{c}  = 0. \label{varSD1standard}
\eeq

By making an infinitesimal  variation along time reparametrization  of   the second Schwinger Dyson equation
 in (\ref{SDyson}) one also finds

 \beq
\delta_{c} \Sigma_{c} = 3 J^2  (G_{c})^2 \delta_{c} G_{c}.
\eeq

that inserted in (\ref{varSD1standard})  gives

\beq
\delta_{c}G_{c} + 3 J^2  G_{c} * (G_{c})^2 \delta_{c} G_{c} *  G_{c}   =  ( 1 -   \tilde{K}_{c} )\delta^{c} G_{c} =  0. \label{zeromodest}
\eeq

\vspace{.5 cm}

In order to compute the effective action in the time reparametrization  one takes non conformal   source term  (\ref{IshiftedReg})

\beq
I_{s} = \int_{0}^{\beta} d \tau d \tau'  G(\tau, \tau') \sigma(\tau, \tau')
 \label{source}
\eeq

and  switch  from $(\tau, \tau')$ coordinates to $(\tau_{+}, \tau_{-})$, slow time and fast time coordinates,   $\tau_{+} = \frac{\tau + \tau'}{2}$, $\tau_{-} = \tau - \tau'$. Then  an expansion   in $\tau_{-}$  is made by keeeping  the leading  $\tau_{-}$ term in the Lagrangian and integrate out $\tau_{-}$. The result is a local effective action in $\tau_{+}$.
 By Taylor expanding  (\ref{soft1})  in powers of $ \tau_{-}$ one finds

 \beq
G(\tau_{+}, \tau_{-}) =  G_{\beta = \infty}(\tau_{-}) \left( 1 + \frac{\Delta}{6} \text{Sch} (f(\tau_{+}), \tau_{+})(\tau - \tau' )^2  + O\left( (\tau - \tau')^4  \right)  \right). \label{expansG}
 \eeq

The  finite temperature version of the rescaled source  $s(\tau, \tau')$ (\ref{sstandard})  along the enhancement direction reads

\beq
s(\tau_1, \tau_2) =  a_0  \frac{\text{sgn}(\tau - \tau')}{\left[J \frac{\beta}{\pi} \sin \left( \frac{\pi}{\beta} |\tau - \tau' | \right)\right]^{2}}U\left(J \log\left(\frac{\beta}{\pi}\sin \left( \frac{\pi}{\beta}| \tau - \tau'|  \right) \right) \right),  \label{sexpression}
\eeq

and   the finite temperature conformal Green function has the form

\beq
G_{c}(\tau, \tau') =  \frac{\text{sgn}(\tau - \tau')}{\left[ J\frac{\beta}{\pi} \sin \left( \frac{\pi}{\beta} |\tau - \tau' | \right)\right]^{1/2}}. \label{thermalcG}
\eeq

By inserting in $I_{s}$, the rescaled source   (\ref{source}),   the leading non trivial order of the  $\tau_{-}$  expansion  (\ref{expansG}),  and   the thermal conformal  Green function  (\ref{thermalcG}) finally gives

\bea
\frac{I_{s}}{N} &=& \frac{a_{0}}{24 \sqrt{J}} \int_{0}^{\beta} d \tau_{+} \text{Sch} (f(\tau_{+}), \tau_{+})
 \int_{-\beta}^{\beta} d \tau_{-} G_{\beta = \infty} (\tau_{-})(J\tau_{-})^{2} \frac{\text{Sgn}(\tau_{-})}{ \left|J\frac{\beta}{\pi} \sin\left( \frac{\pi}{\beta} \tau_{-} \right) \right|^{2 + 1/2}}U\left(\log\left(J|\tau_{-}| \right)  \right) \nn \\
 &\sim& \frac{a_{0}}{24 J}  \int_{0}^{\beta} d \tau_{+}    \text{Sch} (f(\tau_{+}), \tau_{+})            \int_{-\beta}^{\beta}   \frac{ d \tau_{-}}{ J|\tau_{-}| }U\left(\log\left(J|\tau_{-}| \right) \right) \nn \\
 &= &  \frac{\alpha_{S}}{\mathcal{J} }  \int_{0}^{\beta} d \tau \, \,   \text{Sch} (f(\tau ), \tau ).
\eea

In the above expression, the coefficient  $a_0$ has to be fitted numerically. We omitted an overall normalization   $b = \frac{1}{(4\pi)^{1/4}}$ for the SYK  Green functions, which should be included in the definition of the overall coefficients in the last line, (see \cite{Kitaev:2017awl} for full details).

\vspace{.5 cm}

\subsection{Green function  UV response to an enhancing  source}

The path integral for the thermal partition function in the collective field description reads

\beq
Z(\beta) =  \int \mathcal{D} G \mathcal{D} \Sigma e^{N \left( I_{c}(G, \Sigma)  + \int d \tau d \tau' G \sigma  \right) }.
\eeq

The expectation value of the Green function $G(\tau, \tau')$ is given by functional derivative w.r.t. source $\sigma(\tau, \tau')$

\beq
\langle G \rangle = \frac{1}{N} \frac{\delta Z}{\delta \sigma}
\eeq

At large $N$, a good approximation for  $Z$ is given  by steepest descent method. In the strong-coupling/low-temperature
regime $\beta J >> 1$, the saddle points are approximately conformal $(G_c, \Sigma_c )$, and, as illustrated in the previous section of the appendix, one finds by steepest descent

\beq
Z \sim e^{  N I_{c} (G_c, \Sigma_c ) + \frac{N}{12 J^2 | G_c |^2} \langle  \sigma | (1 - \tilde{K}_c )^{-1} \tilde{K}_c | \sigma \rangle + N \langle \sigma | G_c \rangle  }.
\eeq

The above relation gives
\beq
\langle G \rangle  \sim  G_c  +  \frac{1}{6 J^2 | G_c |^2} (1 - \tilde{K}_c )^{-1} \tilde{K}_c | \sigma \rangle,
\eeq

therefore, the non conformal uv contributions to the Green function

\beq
G_{uv} = \langle G \rangle  -   G_c   \sim \frac{1}{6 J^2 | G_c |^2} (1 - \tilde{K}_c )^{-1} \tilde{K}_c | \sigma \rangle \label{Guv}
\eeq

can be computed from the r.h.s. of the previous relation.

   The source has the expansion
\beq
\sigma(\tau, \tau' ) = \sum_{h} a_{h} \frac{ \text{sgn}(\tau - \tau')}{\left|J \frac{\beta}{\pi}\sin\left(\frac{\pi}{\beta} (\tau - \tau')\right)  \right|^{h}}  U\left(J \log\left(\frac{\beta}{\pi}\sin \left( \frac{\pi}{\beta}| \tau - \tau'|  \right) \right) \right).
\eeq

By writing the  smearing functions as a Fourier integral
\beq
 U(\xi) =  \int \frac{d \eta}{2\pi} \tilde{U} (\eta) e^{- i \xi \eta},
\eeq

one has in  (\ref{Guv})

\beq
G_{uv} \sim \frac{1}{6 J^2 | G_c |^2} \sum_{h} a_{h}  \int \frac{d \eta}{2\pi}
\frac{\tilde{k}_{c}(h + i \eta)}{1 - \tilde{k}_{c} (h + i \eta )}
\frac{ \text{sgn}(\tau - \tau')}{|J \frac{\beta}{\pi}\sin \left( \frac{\pi}{\beta}(\tau - \tau') \right) |^{h + i\eta}}  \tilde{U}_{h}(\eta) .
\eeq

Since $U(\xi)$ has support over  $\Delta \xi  \sim \beta J$, the Fourier transform
$\tilde{U}(\eta)$ is non vanishing over $\Delta \eta \sim \frac{2 \pi}{\beta J }$.
In the strong-coupling/low-temperature regime $\beta J >> 1$,
an expansion to the first order in $\eta$ can be employed    $\tilde{k}_{c}(h + i\eta) \sim  \tilde{k}_{c}(h) + i\eta \tilde{k}_{c}'(h)$.
Computation of the eta integral by residues method then  gives

\beq
G_{uv} \sim \frac{i}{6 J^2 | G_c |^2} \sum_{h} a_{h} \tilde{U}(\eta_{h})
 \frac{ \text{sgn}(\tau - \tau')}{| J \frac{\beta}{\pi}\sin \left( \frac{\pi}{\beta}(\tau - \tau') \right) |^{h + i\eta_{h}}},   \qquad   \eta_h =  i \frac{\tilde{k}_{c}(h) - 1}{\tilde{k}'_{c}(h)}. \label{Glast}
\eeq

where   the non vanishing terms in the above sum
are for

\beq
| \eta_h | = \frac{| \tilde{k}_{c}(h) - 1|}{|\tilde{k}_{c}'(h)|} <   \frac{2\pi}{\beta J}.
\eeq

 The Schwarzian term occurs from the  $\eta_h = 0$ term,  for $\tilde{k}_{c}(h) =  1$.  This  occurs for $h=2$ and it is the only relevant term  beyond a certain scale for $\beta J$.
 There is also an intermediate situation where non Schwarzian terms    $\eta_h \ne 0$ contribute to (\ref{Glast}).

To summarize,  in the  $\beta J \rightarrow \infty$ limit, on finds

\beq
G_{UV}(\tau , \tau' ) = \frac{g_{UV}}{|G_{c}|} \sim \alpha_{G}\frac{ \text{sgn}(\tau - \tau')}{| \frac{\beta}{\pi}\sin \left( \frac{\pi}{\beta}(\tau - \tau') \right) |^{3/2}},
\eeq

with

\beq
\alpha_G = \frac{a_{2}}{ k_{c}'(2)}.
\eeq

The $3/2$ exponent is typical of  a quantum system in as  Liouville potential.

\vspace{.5 cm}

\bibliography{SYK}

\providecommand{\href}[2]{#2}\begingroup\raggedright\begin{thebibliography}{10}

\bibitem{Saad:2018bqo}
P.~Saad, S.~H. Shenker and D.~Stanford, \emph{{A semiclassical ramp in SYK and
  in gravity}},  \href{https://arxiv.org/abs/arXiv:1806.06840}{{\ttfamily
  arXiv:1806.06840}}.

\bibitem{Bekenstein:1974ax}
J.~D. Bekenstein, \emph{{Generalized second law of thermodynamics in black hole
  physics}}, \href{https://doi.org/10.1103/PhysRevD.9.3292}{\emph{Phys. Rev.}
  {\bfseries D9} (1974) 3292--3300}.

\bibitem{Hawking:1974sw}
S.~W. Hawking, \emph{{Particle Creation by Black Holes}},
  \href{https://doi.org/10.1007/BF02345020, 10.1007/BF01608497}{\emph{Commun.
  Math. Phys.} {\bfseries 43} (1975) 199--220}.

\bibitem{tHooft:1993dmi}
G.~'t~Hooft, \emph{{Dimensional reduction in quantum gravity}}, {\emph{Conf.
  Proc.} {\bfseries C930308} (1993) 284--296},
  [\href{https://arxiv.org/abs/gr-qc/9310026}{{\ttfamily gr-qc/9310026}}].

\bibitem{Susskind:1993mu}
L.~Susskind and L.~Thorlacius, \emph{{Gedanken experiments involving black
  holes}}, \href{https://doi.org/10.1103/PhysRevD.49.966}{\emph{Phys. Rev.}
  {\bfseries D49} (1994) 966--974},
  [\href{https://arxiv.org/abs/hep-th/9308100}{{\ttfamily hep-th/9308100}}].

\bibitem{Maldacena:1997re}
J.~M. Maldacena, \emph{{The Large N limit of superconformal field theories and
  supergravity}}, \href{https://doi.org/10.1023/A:1026654312961,
  10.4310/ATMP.1998.v2.n2.a1}{\emph{Int. J. Theor. Phys.} {\bfseries 38} (1999)
  1113--1133}, [\href{https://arxiv.org/abs/hep-th/9711200}{{\ttfamily
  hep-th/9711200}}].

\bibitem{Maldacena:2001kr}
J.~M. Maldacena, \emph{{Eternal black holes in anti-de Sitter}},
  \href{https://doi.org/10.1088/1126-6708/2003/04/021}{\emph{JHEP} {\bfseries
  04} (2003) 021}, [\href{https://arxiv.org/abs/hep-th/0106112}{{\ttfamily
  hep-th/0106112}}].

\bibitem{Dyson:2002pf}
L.~Dyson, M.~Kleban and L.~Susskind, \emph{{Disturbing implications of a
  cosmological constant}},
  \href{https://doi.org/10.1088/1126-6708/2002/10/011}{\emph{JHEP} {\bfseries
  10} (2002) 011}, [\href{https://arxiv.org/abs/hep-th/0208013}{{\ttfamily
  hep-th/0208013}}].

\bibitem{Goheer:2002vf}
N.~Goheer, M.~Kleban and L.~Susskind, \emph{{The Trouble with de Sitter
  space}}, \href{https://doi.org/10.1088/1126-6708/2003/07/056}{\emph{JHEP}
  {\bfseries 07} (2003) 056},
  [\href{https://arxiv.org/abs/hep-th/0212209}{{\ttfamily hep-th/0212209}}].

\bibitem{Barbon:2003aq}
J.~L.~F. Barbon and E.~Rabinovici, \emph{{Very long time scales and black hole
  thermal equilibrium}},
  \href{https://doi.org/10.1088/1126-6708/2003/11/047}{\emph{JHEP} {\bfseries
  11} (2003) 047}, [\href{https://arxiv.org/abs/hep-th/0308063}{{\ttfamily
  hep-th/0308063}}].

\bibitem{Horowitz:1999jd}
G.~T. Horowitz and V.~E. Hubeny, \emph{{Quasinormal modes of AdS black holes
  and the approach to thermal equilibrium}},
  \href{https://doi.org/10.1103/PhysRevD.62.024027}{\emph{Phys. Rev.}
  {\bfseries D62} (2000) 024027},
  [\href{https://arxiv.org/abs/hep-th/9909056}{{\ttfamily hep-th/9909056}}].

\bibitem{Papadodimas:2015xma}
K.~Papadodimas and S.~Raju, \emph{{Local Operators in the Eternal Black Hole}},
  \href{https://doi.org/10.1103/PhysRevLett.115.211601}{\emph{Phys. Rev. Lett.}
  {\bfseries 115} (2015) 211601},
  [\href{https://arxiv.org/abs/1502.06692}{{\ttfamily 1502.06692}}].

\bibitem{kitaev:talk1}
A.~Kitaev, \emph{{A simple model of quantum holography (part 1)}},  in
  \emph{KITP Program: Entanglement in Strongly-Correlated Quantum Matter},
  2015,
  \href{http://online.kitp.ucsb.edu/online/entangled15/kitaev/}{http://online.kitp.ucsb.edu/online/entangled15/kitaev/}.

\bibitem{kitaev:talk2}
A.~Kitaev, \emph{{A simple model of quantum holography (part 2)}},  in
  \emph{KITP Program: Entanglement in Strongly-Correlated Quantum Matter},
  2015,
  \href{http://online.kitp.ucsb.edu/online/entangled15/kitaev2/}{http://online.kitp.ucsb.edu/online/entangled15/kitaev2/}.

\bibitem{Sachdev-Ye93}
S.~Sachdev and J.~Ye, \emph{{Gapless spin fluid ground state in a random,
  quantum Heisenberg magnet}},
  \href{https://doi.org/10.1103/PhysRevLett.70.3339}{\emph{Phys. Rev. Lett.}
  {\bfseries 70} (1993) 3339},
  [\href{https://arxiv.org/abs/arXiv:cond-mat/9212030}{{\ttfamily
  arXiv:cond-mat/9212030}}].

\bibitem{Georges}
A.~Georges, O.~Parcollet and S.~Sachdev, \emph{{Mean Field Theory of a Quantum
  Heisenberg Spin Glass}},
  \href{https://doi.org/10.1103/PhysRevLett.85.840}{\emph{Phys. Rev. Lett.}
  {\bfseries 85} (2000) 840},
  [\href{https://arxiv.org/abs/arXiv:cond-mat/9909239}{{\ttfamily
  arXiv:cond-mat/9909239}}].

\bibitem{Sachdev10}
S.~Sachdev, \emph{{Holographic metals and the fractionalized Fermi liquid}},
  \href{https://doi.org/10.1103/PhysRevLett.105.151602}{\emph{Phys. Rev. Lett.}
  {\bfseries 105} (2010) 151602},
  [\href{https://arxiv.org/abs/arXiv:1006.3794}{{\ttfamily arXiv:1006.3794}}].

\bibitem{Sachdev10b}
S.~Sachdev, \emph{{Strange metals and the AdS/CFT correspondence}},
  \href{https://doi.org/10.1088/1742-5468/2010/11/P11022}{\emph{J. Stat. Mech.}
  (2010) 1011}, [\href{https://arxiv.org/abs/arXiv:1010.0682}{{\ttfamily
  arXiv:1010.0682}}].

\bibitem{Sachdev}
S.~Sachdev, \emph{{Bekenstein-Hawking Entropy and Strange Metals}},
  \href{https://doi.org/10.1103/PhysRevX.5.041025}{\emph{Phys. Rev. X}
  {\bfseries 5} (2015) 041025},
  [\href{https://arxiv.org/abs/arXiv:1506.05111}{{\ttfamily
  arXiv:1506.05111}}].

\bibitem{Cotler2017}
J.~S. Cotler, G.~Gur-Ari, M.~Hanada, J.~Polchinski, P.~Saad, S.~H. Shenker
  et~al., \emph{Black holes and random matrices},
  \href{https://doi.org/10.1007/JHEP05(2017)118}{\emph{JHEP} {\bfseries 2017}
  (May, 2017) 118}, [\href{https://arxiv.org/abs/arXiv:1611.04650}{{\ttfamily
  arXiv:1611.04650}}].

\bibitem{Garcia-Garcia:2016mno}
A.~M. García-García and J.~J.~M. Verbaarschot, \emph{{Spectral and
  thermodynamic properties of the Sachdev-Ye-Kitaev model}},
  \href{https://doi.org/10.1103/PhysRevD.94.126010}{\emph{Phys. Rev.}
  {\bfseries D94} (2016) 126010},
  [\href{https://arxiv.org/abs/1610.03816}{{\ttfamily 1610.03816}}].

\bibitem{Khramtsov:2020bvs}
M.~Khramtsov and E.~Lanina, \emph{{Spectral form factor in the double-scaled
  SYK model}},  \href{https://arxiv.org/abs/2011.01906}{{\ttfamily
  2011.01906}}.

\bibitem{Winer:2020gdp}
M.~Winer and B.~Swingle, \emph{{Hydrodynamic Theory of the Connected Spectral
  Form Factor}},  \href{https://arxiv.org/abs/2012.01436}{{\ttfamily
  2012.01436}}.

\bibitem{delCampo:2017bzr}
A.~del Campo, J.~Molina-Vilaplana and J.~Sonner, \emph{{Scrambling the spectral
  form factor: unitarity constraints and exact results}},
  \href{https://doi.org/10.1103/PhysRevD.95.126008}{\emph{Phys. Rev. D}
  {\bfseries 95} (2017) 126008},
  [\href{https://arxiv.org/abs/1702.04350}{{\ttfamily 1702.04350}}].

\bibitem{Cardella:2021rij}
M.~A. Cardella, \emph{{A late times approximation for the SYK spectral form
  factor}},  \href{https://arxiv.org/abs/2102.01653}{{\ttfamily 2102.01653}}.

\bibitem{Witten:2016iux}
E.~Witten, \emph{{An SYK-Like Model Without Disorder}},
  \href{https://arxiv.org/abs/1610.09758}{{\ttfamily 1610.09758}}.

\bibitem{Maldacena:2016hyu}
J.~Maldacena and D.~Stanford, \emph{{Remarks on the Sachdev-Ye-Kitaev model}},
  \href{https://doi.org/10.1103/PhysRevD.94.106002}{\emph{Phys. Rev.}
  {\bfseries D94} (2016) 106002},
  [\href{https://arxiv.org/abs/1604.07818}{{\ttfamily 1604.07818}}].

\bibitem{Jevicki:2016ito}
A.~Jevicki and K.~Suzuki, \emph{{Bi-Local Holography in the SYK Model:
  Perturbations}}, \href{https://doi.org/10.1007/JHEP11(2016)046}{\emph{JHEP}
  {\bfseries 11} (2016) 046},
  [\href{https://arxiv.org/abs/1608.07567}{{\ttfamily 1608.07567}}].

\bibitem{Maldacena:2016upp}
J.~Maldacena, D.~Stanford and Z.~Yang, \emph{{Conformal symmetry and its
  breaking in two dimensional Nearly Anti-de-Sitter space}},
  \href{https://doi.org/10.1093/ptep/ptw124}{\emph{PTEP} {\bfseries 2016}
  (2016) 12C104}, [\href{https://arxiv.org/abs/1606.01857}{{\ttfamily
  1606.01857}}].

\bibitem{Kitaev:2017awl}
A.~Kitaev and S.~J. Suh, \emph{{The soft mode in the Sachdev-Ye-Kitaev model
  and its gravity dual}},
  \href{https://doi.org/10.1007/JHEP05(2018)183}{\emph{JHEP} {\bfseries 05}
  (2018) 183}, [\href{https://arxiv.org/abs/arXiv:1711.08467}{{\ttfamily
  arXiv:1711.08467}}].

\bibitem{Kitaev:2018wpr}
A.~Kitaev and S.~J. Suh, \emph{{Statistical mechanics of a two-dimensional
  black hole}}, \href{https://doi.org/10.1007/JHEP05(2019)198}{\emph{JHEP}
  {\bfseries 05} (2019) 198},
  [\href{https://arxiv.org/abs/1808.07032}{{\ttfamily 1808.07032}}].

\bibitem{Maldacena:2017axo}
J.~Maldacena, D.~Stanford and Z.~Yang, \emph{{Diving into traversable
  wormholes}}, \href{https://doi.org/10.1002/prop.201700034}{\emph{Fortsch.
  Phys.} {\bfseries 65} (2017) 1700034},
  [\href{https://arxiv.org/abs/arXiv:1704.05333}{{\ttfamily
  arXiv:1704.05333}}].

\bibitem{Kourkoulou:2017zaj}
I.~Kourkoulou and J.~Maldacena, \emph{{Pure states in the SYK model and
  nearly-$AdS_2$ gravity}},  \href{https://arxiv.org/abs/1707.02325}{{\ttfamily
  1707.02325}}.

\bibitem{Brustein:2018fkr}
R.~Brustein and Y.~Zigdon, \emph{{Revealing the interior of black holes out of
  equilibrium in the Sachdev-Ye-Kitaev model}},
  \href{https://doi.org/10.1103/PhysRevD.98.066013}{\emph{Phys. Rev.}
  {\bfseries D98} (2018) 066013},
  [\href{https://arxiv.org/abs/1804.09017}{{\ttfamily 1804.09017}}].

\bibitem{Almheiri:2017fbd}
A.~Almheiri, T.~Anous and A.~Lewkowycz, \emph{{Inside out: meet the operators
  inside the horizon. On bulk reconstruction behind causal horizons}},
  \href{https://doi.org/10.1007/JHEP01(2018)028}{\emph{JHEP} {\bfseries 01}
  (2018) 028}, [\href{https://arxiv.org/abs/1707.06622}{{\ttfamily
  1707.06622}}].

\bibitem{Almheiri:2018ijj}
A.~Almheiri, A.~Mousatov and M.~Shyani, \emph{{Escaping the Interiors of Pure
  Boundary-State Black Holes}},
  \href{https://arxiv.org/abs/1803.04434}{{\ttfamily 1803.04434}}.

\bibitem{Almheiri:2018xdw}
A.~Almheiri, \emph{{Holographic Quantum Error Correction and the Projected
  Black Hole Interior}},  \href{https://arxiv.org/abs/1810.02055}{{\ttfamily
  1810.02055}}.

\bibitem{Maldacena:2018lmt}
J.~Maldacena and X.-L. Qi, \emph{{Eternal traversable wormhole}},
  \href{https://arxiv.org/abs/1804.00491}{{\ttfamily 1804.00491}}.

\bibitem{Susskind:2017nto}
L.~Susskind and Y.~Zhao, \emph{{Teleportation through the wormhole}},
  \href{https://doi.org/10.1103/PhysRevD.98.046016}{\emph{Phys. Rev.}
  {\bfseries D98} (2018) 046016},
  [\href{https://arxiv.org/abs/1707.04354}{{\ttfamily 1707.04354}}].

\bibitem{Gao:2016bin}
P.~Gao, D.~L. Jafferis and A.~Wall, \emph{{Traversable Wormholes via a Double
  Trace Deformation}},
  \href{https://doi.org/10.1007/JHEP12(2017)151}{\emph{JHEP} {\bfseries 12}
  (2017) 151}, [\href{https://arxiv.org/abs/1608.05687}{{\ttfamily
  1608.05687}}].

\bibitem{Gu:2018jsv}
Y.~Gu and A.~Kitaev, \emph{{On the relation between the magnitude and exponent
  of OTOCs}}, \href{https://doi.org/10.1007/JHEP02(2019)075}{\emph{JHEP}
  {\bfseries 02} (2019) 075},
  [\href{https://arxiv.org/abs/1812.00120}{{\ttfamily 1812.00120}}].

\bibitem{Maldacena:2015waa}
J.~Maldacena, S.~H. Shenker and D.~Stanford, \emph{{A bound on chaos}},
  \href{https://doi.org/10.1007/JHEP08(2016)106}{\emph{JHEP} {\bfseries 08}
  (2016) 106}, [\href{https://arxiv.org/abs/1503.01409}{{\ttfamily
  1503.01409}}].

\bibitem{Shenker:2013pqa}
S.~H. Shenker and D.~Stanford, \emph{{Black holes and the butterfly effect}},
  \href{https://doi.org/10.1007/JHEP03(2014)067}{\emph{JHEP} {\bfseries 03}
  (2014) 067}, [\href{https://arxiv.org/abs/1306.0622}{{\ttfamily 1306.0622}}].

\bibitem{Dray:1984ha}
T.~Dray and G.~'t~Hooft, \emph{{The Gravitational Shock Wave of a Massless
  Particle}}, \href{https://doi.org/10.1016/0550-3213(85)90525-5}{\emph{Nucl.
  Phys.} {\bfseries B253} (1985) 173--188}.

\bibitem{Sekino:2008he}
Y.~Sekino and L.~Susskind, \emph{{Fast Scramblers}},
  \href{https://doi.org/10.1088/1126-6708/2008/10/065}{\emph{JHEP} {\bfseries
  10} (2008) 065}, [\href{https://arxiv.org/abs/0808.2096}{{\ttfamily
  0808.2096}}].

\bibitem{Hayden:2007cs}
P.~Hayden and J.~Preskill, \emph{{Black holes as mirrors: Quantum information
  in random subsystems}},
  \href{https://doi.org/10.1088/1126-6708/2007/09/120}{\emph{JHEP} {\bfseries
  09} (2007) 120}, [\href{https://arxiv.org/abs/0708.4025}{{\ttfamily
  0708.4025}}].

\bibitem{Brown:2016wib}
A.~R. Brown, L.~Susskind and Y.~Zhao, \emph{{Quantum Complexity and Negative
  Curvature}}, \href{https://doi.org/10.1103/PhysRevD.95.045010}{\emph{Phys.
  Rev.} {\bfseries D95} (2017) 045010},
  [\href{https://arxiv.org/abs/1608.02612}{{\ttfamily 1608.02612}}].

\bibitem{Lin:2019qwu}
H.~W. Lin, J.~Maldacena and Y.~Zhao, \emph{{Symmetries Near the Horizon}},
  \href{https://arxiv.org/abs/1904.12820}{{\ttfamily 1904.12820}}.

\bibitem{Saad:2019lba}
P.~Saad, S.~H. Shenker and D.~Stanford, \emph{{JT gravity as a matrix
  integral}},  \href{https://arxiv.org/abs/1903.11115}{{\ttfamily 1903.11115}}.

\bibitem{Teitelboim:1983ux}
C.~Teitelboim, \emph{{Gravitation and Hamiltonian Structure in Two Space-Time
  Dimensions}}, \href{https://doi.org/10.1016/0370-2693(83)90012-6}{\emph{Phys.
  Lett.} {\bfseries 126B} (1983) 41--45}.

\bibitem{Jackiw:1984je}
R.~Jackiw, \emph{{Lower Dimensional Gravity}},
  \href{https://doi.org/10.1016/0550-3213(85)90448-1}{\emph{Nucl. Phys.}
  {\bfseries B252} (1985) 343--356}.

\bibitem{Stanford:2019vob}
D.~Stanford and E.~Witten, \emph{{JT Gravity and the Ensembles of Random Matrix
  Theory}},  \href{https://arxiv.org/abs/1907.03363}{{\ttfamily 1907.03363}}.

\bibitem{Maldacena:2013xja}
J.~Maldacena and L.~Susskind, \emph{{Cool horizons for entangled black holes}},
  \href{https://doi.org/10.1002/prop.201300020}{\emph{Fortsch. Phys.}
  {\bfseries 61} (2013) 781--811},
  [\href{https://arxiv.org/abs/1306.0533}{{\ttfamily 1306.0533}}].

\bibitem{Israel:1976ur}
W.~Israel, \emph{{Thermo field dynamics of black holes}},
  \href{https://doi.org/10.1016/0375-9601(76)90178-X}{\emph{Phys. Lett.}
  {\bfseries A57} (1976) 107--110}.

\bibitem{DS}
D.~Stanford, \emph{Private communication}, .

\bibitem{Zhang:2020jhn}
P.~Zhang, Y.~Gu and A.~Kitaev, \emph{{An obstacle to sub-AdS holography for
  SYK-like models}},  \href{https://arxiv.org/abs/2012.01620}{{\ttfamily
  2012.01620}}.

\bibitem{Sarosi:2017ykf}
G.~Sárosi, \emph{{AdS$_{2}$ holography and the SYK model}},
  \href{https://doi.org/10.22323/1.323.0001}{\emph{PoS} {\bfseries Modave2017}
  (2018) 001}, [\href{https://arxiv.org/abs/1711.08482}{{\ttfamily
  1711.08482}}].

\bibitem{Stanford:2017thb}
D.~Stanford and E.~Witten, \emph{{Fermionic Localization of the Schwarzian
  Theory}}, \href{https://doi.org/10.1007/JHEP10(2017)008}{\emph{JHEP}
  {\bfseries 10} (2017) 008},
  [\href{https://arxiv.org/abs/1703.04612}{{\ttfamily 1703.04612}}].

\bibitem{Bagrets:2016cdf}
D.~Bagrets, A.~Altland and A.~Kamenev, \emph{{Sachdev–Ye–Kitaev model as
  Liouville quantum mechanics}},
  \href{https://doi.org/10.1016/j.nuclphysb.2016.08.002}{\emph{Nucl. Phys.}
  {\bfseries B911} (2016) 191--205},
  [\href{https://arxiv.org/abs/1607.00694}{{\ttfamily 1607.00694}}].

\bibitem{Bagrets:2017pwq}
D.~Bagrets, A.~Altland and A.~Kamenev, \emph{{Power-law out of time order
  correlation functions in the SYK model}},
  \href{https://doi.org/10.1016/j.nuclphysb.2017.06.012}{\emph{Nucl. Phys.}
  {\bfseries B921} (2017) 727--752},
  [\href{https://arxiv.org/abs/1702.08902}{{\ttfamily 1702.08902}}].

\bibitem{Mertens:2017mtv}
T.~G. Mertens, G.~J. Turiaci and H.~L. Verlinde, \emph{{Solving the Schwarzian
  via the Conformal Bootstrap}},
  \href{https://doi.org/10.1007/JHEP08(2017)136}{\emph{JHEP} {\bfseries 08}
  (2017) 136}, [\href{https://arxiv.org/abs/arXiv:1705.08408}{{\ttfamily
  arXiv:1705.08408}}].

\bibitem{Mertens:2018fds}
T.~G. Mertens, \emph{{The Schwarzian Theory - Origins}},
  \href{https://doi.org/10.1007/JHEP05(2018)036}{\emph{JHEP} {\bfseries 05}
  (2018) 036}, [\href{https://arxiv.org/abs/1801.09605}{{\ttfamily
  1801.09605}}].

\bibitem{Harlow:2018tqv}
D.~Harlow and D.~Jafferis, \emph{{The Factorization Problem in
  Jackiw-Teitelboim Gravity}},
  \href{https://arxiv.org/abs/1804.01081}{{\ttfamily 1804.01081}}.

\bibitem{Blommaert:2018iqz}
A.~Blommaert, T.~G. Mertens and H.~Verschelde, \emph{{Fine Structure of
  Jackiw-Teitelboim Quantum Gravity}},
  \href{https://arxiv.org/abs/1812.00918}{{\ttfamily 1812.00918}}.

\end{thebibliography}\endgroup
\bibliographystyle{JHEP}
\end{document}